\documentclass[twocolumn]{aastex63}

\usepackage{lipsum, babel}

\newcommand\tess{TESS}

\newcommand{\PSUAA}{Department of Astronomy \& Astrophysics, 525 Davey Laboratory, Penn State, University Park, PA, 16802, USA}
\newcommand{\PSUCEHW}{Center for Exoplanets and Habitable Worlds, 525 Davey Laboratory, Penn State, University Park, PA, 16802, USA}
\newcommand{\PSETI}{Penn State Extraterrestrial Intelligence Center, 525 Davey Laboratory, Penn State, University Park, PA, 16802, USA}
\newcommand{\UA}{Steward Observatory, University of Arizona, 933 N.\ Cherry Ave, Tucson, AZ 85721, USA}

\newcommand{\Penn}{Department of Physics and Astronomy, University of Pennsylvania, 209 S 33rd St, Philadelphia, PA 19104, USA}

\newcommand{\GoddardESAL}{Exoplanets and Stellar Astrophysics Laboratory, NASA Goddard Space Flight Center, Greenbelt, MD 20771, USA}

\newcommand{\NOAO}{NSF National Optical-Infrared Astronomy Research Laboratory, 950 N.\ Cherry Ave., Tucson, AZ 85719, USA}

\newcommand{\Macquarie}{School of Mathematical and Physical Sciences, Macquarie University, Balaclava Road, North Ryde, NSW 2109, Australia}

\newcommand{\CUBoulder}{Department of Physics, 390 UCB, University of Colorado, Boulder, CO 80309, USA}
\newcommand{\JPL}{Jet Propulsion Laboratory, California Institute of Technology, 4800 Oak Grove Drive, Pasadena, California 91109}

\newcommand{\UCI}{Department of Physics \& Astronomy, The University of California, Irvine, Irvine, CA 92697, USA}
\newcommand{\Carleton}{Carleton College, One North College St., Northfield, MN 55057, USA}

\newcommand{\UCO}{The University of California Observatories, 1156 High Street, Santa Cruz, CA 95064, USA}

\newcommand{\TIFR}{Department of Astronomy and Astrophysics, Tata Institute of Fundamental Research, Homi Bhabha Road, Colaba, Mumbai 400005, India}

\newcommand{\UAm}{Anton Pannekoek Institute for Astronomy, 904 Science Park, University of Amsterdam, Amsterdam, 1098 XH, The Netherlands}

\submitjournal{ApJ}

\begin{document}

\title{Jitter Across 15 Years: Leveraging Precise Photometry from Kepler and TESS to Extract Exoplanets from Radial Velocity Time Series}

\author[0000-0001-7708-2364]{Corey Beard}
\altaffiliation{NASA FINESST Fellow}
\affiliation{\UCI}

\author[0000-0003-0149-9678]{Paul Robertson}
\affiliation{\UCI}

\author[0000-0001-8342-7736]{Jack Lubin}
\affiliation{Department of Physics \& Astronomy, University of California Los Angeles, Los Angeles, CA 90095, USA}

\author[0000-0002-7127-7643]{Te Han}
\affiliation{\UCI}

\author[0000-0002-5034-9476]{Rae Holcomb}
\affiliation{\UCI}

\author[0000-0001-5728-4735]{Pranav Premnath}
\affiliation{\UCI}


\author[0000-0003-1305-3761]{R. Paul Butler}
\affiliation{Earth and Planets Laboratory, Carnegie Institution for Science, Washington, DC 20015, USA}

\author[0000-0002-4297-5506]{Paul A.\ Dalba}
\altaffiliation{Heising-Simons 51 Pegasi b Postdoctoral Fellow}
\affiliation{Department of Astronomy and Astrophysics, University of California, Santa Cruz, CA 95064, USA}
\affiliation{SETI Institute, Carl Sagan Center, 339 Bernardo Ave, Suite 200, Mountain View, CA 94043, USA}

\author[0000-0002-6153-3076]{Brad Holden}
\affiliation{\UCO}


\author[0000-0002-6096-1749]{Cullen H.\ Blake}
\affil{\Penn}

\author[0000-0002-2144-0764]{Scott A.\ Diddams}
\affil{\CUBoulder}

\author[0000-0002-5463-9980]{Arvind F.\ Gupta}
\affil{\NOAO}

\author[0000-0003-1312-9391]{Samuel Halverson}
\affiliation{\JPL}

\author[0000-0001-9626-0613]{Daniel M.\ Krolikowski}
\affil{\UA}

\author[0000-0001-7318-6318]{Dan Li}
\affil{\NOAO}

\author[0000-0002-9082-6337]{Andrea S.J.\ Lin}
\affil{\PSUAA}
\affil{\PSUCEHW}

\author[0000-0002-9632-9382]{Sarah E.\ Logsdon}
\affil{\NOAO}

\author[0000-0003-0790-7492]{Emily Lubar}
\affil{Aerospace Corporation, 200 N Aviation Blvd, El Segundo, CA, 90245, USA}

\author[0000-0001-9596-7983]{Suvrath Mahadevan}
\affil{\PSUAA}
\affil{\PSUCEHW}

\author[0000-0003-0241-8956]{Michael W.\ McElwain}
\affil{\GoddardESAL} 

\author[0000-0001-8720-5612]{Joe P.\ Ninan}
\affil{\TIFR}

\author[0000-0003-1324-0495]{Leonardo A.\ Paredes}
\affil{\UA}

\author[0000-0001-8127-5775]{Arpita Roy}
\affiliation{Astrophysics \& Space Institute, Schmidt Sciences, New York, NY 10011, USA}

\author[0000-0002-4046-987X]{Christian Schwab}
\affil{\Macquarie}

\author[0000-0001-7409-5688]{Gudmundur Stefansson}
\affil{\UAm}

\author[0000-0002-4788-8858]{Ryan C. Terrien}
\affil{\Carleton}

\author[0000-0001-6160-5888]{Jason T.\ Wright}
\affil{\PSUAA}
\affil{\PSUCEHW}
\affil{\PSETI}

\correspondingauthor{Corey Beard}
\email{ccbeard@uci.edu}

\begin{abstract}

Stellar activity contamination of radial velocity (RV) data is one of the top challenges plaguing the field of extreme precision RV (EPRV) science. Previous work has shown that photometry can be very effective at removing such signals from RV data, especially stellar activity caused by rotating star spots and plage.The exact utility of photometry for removing RV activity contamination, and the best way to apply it, is not well known. We present a combination photometric and RV study of eight Kepler/K2 FGK stars with known stellar variability. We use NEID RVs acquired simultaneously with TESS photometry, and we perform injection recovery tests to quantify the efficacy of recent TESS photometry versus archival Kepler/K2 photometry for removing stellar variability from RVs. We additionally experiment with different TESS sectors when training our models in order to quantify the real benefit of simultaneously acquired RVs and photometry. We conclude that Kepler photometry typically performs better than TESS at removing noise from RV data when it is available, likely due to longer baseline and precision. In contrast, for targets with available K2 photometry, especially those most active, and with high precision ($\sigma_{NEID}$ $<$ 1 m s$^{-1}$) NEID RVs, TESS may be the more informative dataset. However, contrary to expectations, we have found that training on simultaneous photometry does not always achieve the best results.

\end{abstract}


\section{Introduction} \label{sec:intro}

Radial velocity (RV) analysis of stars is one of the oldest \citep{mayor95} and most successful \citep{lagrange09,reiners18,rosenthal21} methods for discovering and characterizing exoplanets. RV observations additionally provide an excellent means of validating transiting exoplanets \citep[e.g.][]{plavchan20,beard22_A,beard22_b}, providing mass measurements \citep{haywood14,LM16, rajpaul21}, and probing regions of parameter space that transit observations rarely can \citep[i.e. long-period, distant planets][]{lubin22}. Measuring exoplanet masses is especially relevant in the era of the James Webb Space Telescope \citep[JWST;][]{gardner06}, as a precise mass measurement is necessary to interpret atmospheric transmission spectra \citep{batalha19}. Further, the discovery of exoplanets around the nearest stars is an essential precursor to a future Habitable Worlds Observatory \citep[HWO;][]{decadal20,mamajek24} mission to image Habitable Zone \citep[HZ;][]{kasting93,kopparapu13} planets around the nearest stars, and the vast majority such planets are unlikely to transit \citep{hardegree23}. To maximize the efficiency of future missions such as HWO, a curated list of known, nearby planets is required, and RV detection remains our only plausible method for discovering such nearby, imagable exoplanets.

RV exoplanet science faces serious challenges, however, especially in the era of extreme precision RV (EPRV) observations \citep{wright17}. As we push instrument stability below the historical 1 m s$^{-1}$ noise floor,  a variety of physical processes in stellar atmospheres contaminate our RV data at amplitudes larger than the instrumental precision, making it difficult to identify low-amplitude exoplanets \citep[e.g.][]{LM16,blunt23, beard24}. This contamination can vary widely in amplitude and frequency. Cool spots or hot plage on the surfaces of stars can quasi-periodically deform stellar spectra, producing undesired red or blue shifts in our spectra that do not originate from gravitational reflex, having effects up to 1000+ m s$^{-1}$ \citep{saar97,meunier10}. Granulation and stellar p-mode oscillations are another class of contamination that can have effects up to 10 m s$^{-1}$ \citep{chaplin19}.

A variety of methods exist for modeling out or otherwise accounting for stellar contamination of RV data, varying in complexity and effectiveness. Adding a ``jitter" term in quadrature with RV uncertainties is perhaps the simplest method, though ineffective in cases where noise is correlated and large in amplitude. Another common technique is decorrelating RVs to some stellar activity indicator, metrics extracted from spectra to track the activity of the host star \citep{robertson16,lafarga20}. Unfortunately, the correlation between RV activity contamination and activity indicators is not always well-governed by a simple relationship \citep[i.e. time delays between the datasets][]{burrows24}. Further, many analyses today utilize data from multiple instruments, and it is common that different instruments do not track the same activity indicators. Gaussian Process \citep[GP;][]{ambikasaran15} regression is one of the most common and effective ways to remove the effects of stellar magnetic activity from RV data \citep{haywood14, LM16, beard24}, though how exactly to utilize it is not always clear. GPs are flexible and non-physical, and imposing bounds on their flexibility can be extremely helpful for separating quasi-periodic stellar variability from true exoplanet signals. GPs can be trained on stellar activity indicators, or fit with them simultaneously \citep{rajpaul15, rajpaul21}. Utilizing photometry is another popular method for removing stellar variability from RV data \citep{aigrain12,grunblatt15,tran23}, especially if taken contemporaneously. Photometric data, unlike spectroscopic stellar activity indicators, are often much higher in cadence and precision, allowing for a better characterization of the current stellar astrophysics. Additionally, large photometric datasets exist for many systems, and photometry can often be acquired for thousands of targets simultaneously. Astronomers generally agree that simultaneous photometry is a powerful tool when mitigating stellar variability in RVs, but how powerful? Multiple photometric datasets exist, with different precisions and observing baselines. Are there clear reasons to use one over another, or is utilizing all available photometry the best course?

In this work we seek to study the effectiveness of photometric datasets when used to correct for stellar variability in RVs. We also wish to understand the importance of simultaneous, or near-simultaneous photometry and RV data. Doing so will require answering several key questions.  Firstly, which stellar properties are constrained by which lightcurves? TESS's 27-day baseline for most stars suggests that TESS might be superior for constraining shorter-lived activity, while Kepler's long baseline makes it ideal for longer activity cycles.

Secondly, which lightcurves should be used to correct for stellar activity? RV noise may be suppressed using a joint model to data from Kepler, TESS, or both. Most likely, this will depend upon the temporal proximity of RVs and photometry.

Finally, how well can photometry remove stellar activity contamination from RVs. Newer instruments such as NEID are capable of achieving RV precision better than 30 cm s$^{-1}$ \citep{schwab16} for bright targets, allowing us to constrain photometric variability and Doppler jitter more precisely than was possible for Kepler systems. The most complete analysis relating Kepler photometric variability to RV jitter was conducted using Keck/HIRES, which has an instrumental noise floor of 2-3 m s$^{-1}$ \citep{bastien14}.

To gain insights into these questions, we chose a variety of bright targets with either Kepler or K2 photometry, and we observed them with NEID and TESS simultaneously. We detail our selection criteria, targets, and their stellar parameters in \S \ref{sec:target_selection}, the data we use in \S \ref{sec:data}, our primary analysis in \S \ref{sec:analysis}, and we discuss the results in \S \ref{sec:discussion}.

\section{Target Selection}
\label{sec:target_selection}

We elected to begin a study of a variety of bright targets in order to better understand the role photometry plays in stellar activity mitigation. Over the course of this study, we chose targets with particular features. Targets were chosen for the express purpose of studying photometry's ability to mitigate stellar variability in RVs. Two criteria were mandatory: first, we chose targets that had existing Kepler or K2 photometry. Comparing the value of TESS training versus older Kepler/K2 training forms a core part of this analysis, and could not be done without such archival data. Second, we were interested in achieving simultaneous NEID observations during TESS observations. Consequently, when choosing targets for a NEID observing semester, we required that they be observed by TESS at a time that NEID, too, could observe them.

We then chose a number of other optional, but desirable, features. We wanted bright targets to minimize the amount of telescope time required to observe our targets at high precision. We imposed a general magnitude cut of V $<$ 10. We also chose targets that had archival HIRES, HARPS, or HARPS-N RVs. This allowed us to 1) confirm that these targets were amenable to precise RV observations, and 2) approximate to first order the level of activity contamination that would likely exist in the RV data. 

We mainly restricted our observations to main sequence FGK stars, for a number of reasons. Firstly, these were the most abundant stars observed during the Kepler mission. Secondly, more evolved stars begin to see non-spot dominated forms of stellar activity emerge as highly significant, such as p-mode oscillations and granulation. These generally reduce our RV precision and interfere with our analysis, which is primarily focused on spot modulation. Finally, M dwarf targets typically require a different treatment. For example,  \cite{aigrain12}'s FF$^{\prime}$ method assumes a single large spot on the star, which is not a good approximation for M Dwarf photospheres. Furthermore, stellar activity can function very distinctly from that on FGK dwarfs and should not be lumped into a single analysis \citep{hojjatpanah20}.

The above restrictions left us with a pool of targets to choose from. We prioritized the brightest, with special weight given to those that exhibited spot-induced stellar variability in their Kepler or K2 photometry. We discuss our eight targets briefly in the next subsections.

\subsubsection{HD 173701}

HD 173701 is a bright Kepler star with no known exoplanets. Because it is a solar analogue in age, radius, and mass, it has seen previous study \citep{thomas19}. The star is known to exhibit differential rotation, with rotation rates at 45$^{o}$ latitude $\sim$ 50$\%$ slower than at the equator \citep{benomar18}. and a number of archival HIRES RVs have been taken to study the star \citep{rosenthal21}. We additionally obtained new APF RVs of HD 173701 and use them as as part of our analysis.

A generalized Lomb-Scargle \citep[GLS;][]{zechmeister18} periodogram on HD 173701 Kepler photometry reveals high powered signals between 32 and 36 days, suggesting a possible rotation period somewhere in this range, though the system is known to differentially rotate. The same analysis on TESS photometry highlights a strong signal near 12 days, most likely a harmonic of the longer period signal identified in Kepler. TESS is generally insensitive to signals near or longer than half a TESS sector in length ($\sim$12 days), though TESS can sometimes identify physical signals at the top end of this range \citep{holcomb22}. The system likely has a rotation period near 32 days, and TESS only detects a harmonic.

\subsubsection{Kepler-21}

Kepler-21 is the brightest known Kepler system with a transiting planet, and has a variety of scientifically interesting features. The system was first studied in \cite{howell12}, where a small, transiting exoplanet was discovered orbiting the star with a period near 2.7 days. \cite{LM16} performed a joint photometric-RV analysis of the system, and determined the composition of the planet to be consistent with that of Earth. Their mass estimate, however, was hindered by stellar variability, as the system is known to be contaminated with stellar activity (K$_{b}$ = 2.46$\pm$0.48 m s$^{-1}$; RV RMS = 4.95 m s$^{-1}$). More recently, \cite{bonomo23} obtained more data and significantly increased our confidence in the planet mass measurement, placing it near 7.5 $\pm$ 1.3 M$_{\oplus}$, and consistent with an Earth-like density. We perform a detailed study of Kepler-21 during a precursor analysis (Beard et al. 2024b, submitted), though we summarize the system briefly here.

Kepler-21 is actually a slightly evolved F4-6 IV star (6250 $\pm$ 250 K). Despite this fact, we still include it in our observations because it is historically important to the community, is very bright, has a known rotation period $\sim$ 12.7 days, and has an abundance of archival RVs.

\subsubsection{Kepler-37}

Kepler-37 is the dimmest star in our sample  \citep[V = 9.77$\pm$0.03;][]{rajpaul21}, but still relatively bright. The system has three known transiting exoplanets, with a fourth controversial candidate \citep{rajpaul21}. Most remarkable about the system is the size of its exoplanets, with Kepler-37 b smaller than Mercury, approximately the size of the Earth's Moon \citep{barclay13}. Planets b and c have RV amplitudes far too small for study with even the highest precision instruments today (K$_{b}$ $<$ 1 cm s$^{-1}$, K$_{c}$ $<$ 14 cm s$^{-1}$), though \cite{rajpaul21} were able to constrain the RV signal of Kepler-37 d (1.22$\pm$0.31 m s$^{-1}$), one of the smallest ever detected RV amplitudes at the time.

The Kepler photometry of Kepler-37 is contaminated with a quasi-periodic signal suggestive of spot modulation, and the system has a likely rotation period of 29 days. Consequently, Kepler-37 makes an interesting test case to compare activity training between Kepler and TESS, especially with a suspected rotation period longer than the baseline of a single TESS sector. All three transiting exoplanets have small RV semi-amplitudes, though Kepler-37 d is plausibly detectable with NEID, and was detected using HARPS-N. Consequently, when modeling Kepler-37 in our analysis, we treat it as a one planet system containing Kepler-37 d, and ignore the other two planets, far below our sensitivity.

\subsubsection{HD 4256}

HD 4256 is a bright (V = 8.00) K3 dwarf with no known exoplanets that was observed during the K2 mission. It has a long history of HIRES observations \citep{rosenthal21}, and a clear, long period, periodic signal is apparent from the RVs. This signal is strongly correlated with the Calcium II H\&K S index (S$_{HK}$), however, suggesting this cycle is not planetary, but likely related to stellar variability. After subtracting this signal, modest residual scatter remained, suggesting planets or additional stellar variability. The presence of the long-term activity cycle, clearly seen in S$_{HK}$ values, motivated us to test if, in fact, such a correlation existed for photometry as well.

\subsubsection{HD 31966}

HD 31966 is a bright G2 dwarf that saw observations during the K2 mission. The system has no known exoplanets, though it has been considered as a promising target for asteroseismology \citep{schofield19}. The target was chosen for our project mainly because it is extremely bright, has archival HIRES RVs \citep{rosenthal21}, and we detected scatter in the RVs suggestive of either stellar variability, or an undetected exoplanet. 

\subsubsection{HD 24040}

HD 24040 is a bright (V = 7.51) G1 star with a long history of HIRES RV observations. The system has two known RV-detected exoplanets with very long periods, one larger than Jupiter, and the other near Saturn in size. The system also has a long term linear trend that has begun to show signs of curvature \citep{rosenthal21}. The trend/curvature is of sufficiently low amplitude that it may well be a long period planet. 

The system was also observed during the K2 mission, making it a potentially fruitful target for study. Often, when selecting targets, we would look at the archival Kepler or K2 photometry to get a first order estimate of stellar activity contamination. This is somewhat difficult in K2 photometry, due to K2's ubiquitous systematic contamination issues (mentioned further in \S \ref{sec:data}), though an analysis of archival RV data suggests that after subtracting the three known signals in the RVs, there was a few m s$^{-1}$ of residual scatter, suggesting either stellar variability or additional planets. We are interested in exploring activity near the historical 1 m s$^{-1}$ noise floor in addition to more obvious spot modulation, and so we included HD 24040 in our target list.

\subsubsection{HD 106315}

HD 106315 is a bright (V=8.95) system with two known sub-Neptune transiting exoplanets \citep{crossfield17} that was observed by K2. \citet{barros17} measured the masses of the transiting exoplanets using HARPS radial velocities. The system's host star is an F5 star with a suspected rotation period near 5 days. Due to the rapid rotation and early spectral type of the star, line broadening typically reduces the precision of RV observations. Nonetheless, due to its brightness, HD 106315 b and c are excellent targets for atmospheric observations \citep{kreidberg22}.

After systematic correction, K2 photometry exhibit a clear quasi-periodic signal, likely due to spot modulation, making it a useful addition to our study. 

\subsubsection{HD 119291}

HD 119291 is a star with no known transiting exoplanets that was observed by K2. The target is very bright \citep[9.24$\pm$0.03;][]{stassun19}, and we observed that the K2 photometry exhibits distinctive quasi-periodic modulation that is likely due to spot modulation. HD 119291 has also seen archival observations with the HARPS spectrograph, mainly with respect to Gaia radial velocity standard stars \citep{soubiran18}.

\subsection{Stellar Parameters}

As alluded to previously, all of our targets are very bright, and have characterized in previous works. Rather than repeat such analyses, we summarize the stellar parameters of our targets in Table \ref{tab:stellar}.  While such a heterogeneous collection of stellar properties is unsuitable for demographic studies of stellar or exoplanetary populations, it is sufficient to describe the basic properties of targets for our study, which is primarily focused on time-series analysis.

\begin{deluxetable*}{cccccccccc}
\label{tab:stellar}
\tablecaption{Stellar Parameters}
\tablehead{\colhead{System}  &  \colhead{T$_{\rm{eff}}$ (K)}
& \colhead{Spectral Type} & \colhead{R$_{*}$ (R$_{\odot}$)}  & \colhead{M$_{*}$ (M$_{\odot}$)}   & \colhead{L$_{*}$ (L$_{\odot}$)} & \colhead{V Magnitude}  & \colhead{$\log \rm{R}^{\prime}\rm{HK}$} & P$_{pred}$ (days) & \colhead{Reference}}
\startdata
HD 173701 & 5337$\pm$105 & G8V & 0.96$\pm$0.04 & 0.92$\pm$0.11 & 0.67$\pm$0.01 & 7.54$\pm$0.03 & -4.94 & 41.38 & A \\
 Kepler-21 & 6305$\pm$50 & F6IV & 1.902$^{+0.018}_{-0.012}$ & 1.408$^{+0.021}_{-0.030}$& 5.188$^{+0.142}_{-0.128}$ & 8.25$\pm$0.03 & -5.19 & 14.83 & B \\
Kepler-37 & 5406$\pm$28 & G8V & 0.787$^{+0.033}_{-0.031}$ & 0.87$\pm$0.15 & 0.479$\pm$0.001 & 9.77$\pm$0.03 & -4.93 & 26.5 & C  \\
HD 4256 & 5017$\pm$141 & K3V & 0.77$\pm$0.06 & 0.83$\pm$0.11 & 0.33$\pm$0.01 & 8.045$\pm$0.013 & -4.95 & 46.2 & A \\
 HD 31966 & 5715$\pm$108  & G2IV-V & 1.61$\pm$0.07 & 1.02$\pm$0.12 & 2.48$\pm$0.06 & 6.74$\pm$0.02 & -5.06 & 29.0 & A \\
 HD 24040 & 5776$\pm$84 & G1V & 1.38$\pm$0.03 & 1.10$\pm$0.05 & 0.27$\pm$0.01 & 7.515$\pm$0.009 & -5.05 & 27.7 & D \\
 HD 106315 & 6321$\pm$50 & F5V & 1.27$^{+0.17}_{-0.13}$ & 1.12$^{+0.05}_{-0.04}$ & 0.388$\pm$0.004 & 8.951$\pm$0.003 & -5.14 & 5.9 & E \\
 HD 119291 & 4510$\pm$137 & K7V & 0.69$\pm$0.06 & 0.71$\pm$0.08 & 0.178$\pm$0.009 & 9.24$\pm$0.03 & -4.95 & 46.2 & F \\
\enddata
\tablenotetext{}{A is \cite{stassun19}, B is \cite{LM16}, C is \cite{rajpaul21}, D is \cite{rosenthal21}, E is \cite{mayo18}, and F is \cite{soubiran18}. }
\end{deluxetable*}

We predict rotation periods using an estimate from the $\log R^{\prime}_{HK}$ activity indicators given in \cite{noyes84}. We call these $P_{\rm{pred}}$ in Table \ref{tab:stellar}. A few of our targets have known rotation periods (Kepler-21: 12.6$\pm$0.03 days \citep{LM16}; Kepler-37: 29$\pm$1 days \citep{rajpaul21}; HD 106315: 5.15$\pm$0.28 days \citep{barros17}) and most of the others show clear signs of periodic modulation in photometry and RVs. Photometry can often produce more reliable estimates of stellar rotation periods \citep{mcquillan14,holcomb22}, though we utilize a non-photometric method to prevent ``double fitting" photometry, which we use later in \S \ref{sec:analysis}. We have found that the estimate in \cite{noyes84} is generally close to the known or suspected rotation periods of our targets. We note that Kepler-21's predicted rotation period using \cite{noyes84} was 0.05 days, far from the known, true value. This might be caused by the relationship in \cite{noyes84} failing for a more evolved star. Instead, we use the asteroseismological estimate from \cite{howell12} for our predicted rotation estimate for Kepler-21.

As mentioned above, we observe primarily main sequence FGK stars. The primary exception to this rule is Kepler-21, as well as the slightly evolved HD 31966. We adopt stellar parameters from planet discovery papers for those with known planets, and use the TICv8 catalog for the remainder \citep{stassun19}, as well as to supplement missing parameters when not included in other papers.

The observed photometric variability in our sample is indicative of spot dominated stellar activity in general \citep{dumusque14}, though some other forms of stellar activity are likely present in the data at lower amplitudes.

\section{Survey Data}
\label{sec:data}

\subsection{Photometric Data}

We chose targets with an abundance of photometric data. Kepler targets that were able to be observed simultaneously with NEID and TESS were prioritized, but TESS pointing constraints forced us to choose several targets with K2 photometry instead of Kepler.

\subsubsection{Kepler Photometry}

Three of our targets were Kepler targets, meaning that they were observed as a part of the primary Kepler mission that launched on 6 March 2009, with observations beginning 2 May 2009. All three of these targets--Kepler-21, HD 173701, and Kepler-37--were observed until 11 May 2013, spanning 1470 days \citep{borucki10}. The Kepler spacecraft utilized a 1.4 m primary mirror to observe $\sim$ 150,000 main sequence stars using its 115
square degree field of view.

Kepler observations were divided into ``quarters," 90 days in length. Both short and long cadence observations were taken by Kepler, with exposure times of 58.85 s and 29.4 minutes respectively. Long cadence data is available for all Kepler quarters, while short cadence data is only available for quarters 2, and 5-17. We choose to use long cadence data during our analysis, as our RV cadence is insensitive to activity on sub-hour timescales.

We utilize the Presearch Data Conditioning (PDC) flux, produced by the Kepler science processing pipeline \citep[KSPP;][]{jenkins10}. This pipeline reduces raw data into a processed form, removing known instrumental and erroneous effects, and flagging datapoints of suspicious quality. Reductions can remove genuine physical signals from the photometry on periods near the length of a Kepler quarter (90 days), though our targets are either known to, or suspected to, have rotation periods far less than this length. A summary of our Kepler data is visible in Table \ref{tab:observations}.

\begin{deluxetable*}{lcccccccc}
\label{tab:observations}
\tablecaption{Observational Statistics}
\tablehead{\colhead{Instrument}  &  \colhead{HD 173701}
& \colhead{Kepler-21} & \colhead{Kepler-37}  & \colhead{HD 4256}   & \colhead{HD 31966}   & \colhead{ HD 24040}   & \colhead{HD 106315}   & \colhead{HD 119291}}
\startdata
\textbf{Photometric Data} \\
~~~~ Median Kepler Precision & 2.92 ppm & 4.38 ppm & 7.79 ppm & - & - & - & - & - \\
 ~~~~ Median K2 Precision & - & - & - & 3.77 ppm & 4.83 ppm & 6.50 ppm & 13.14 ppm & 13.10 ppm \\
 ~~~~ Median TESS Precision & 23.89 ppm & 38.83 ppm  & 79.62 ppm & 27.44 ppm & 17.27 ppm & 24.64 ppm & 489 ppm & 61.01 ppm \\
\textbf{HIRES-pre Data} \\
~~~~ Median Precision & - & - & - & 1.3 m s$^{-1}$ & 1.27 m s$^{-1}$ & 1.3 m s$^{-1}$ & - & - \\
~~~~ N$_{RV}$ & - & - & - & 27 & 16 & 21 & - & - \\
~~~~ RV baseline & - & - & - & 2829 days & 2059 days & 2369 days & - & - \\
~~~~ Median Cadence & - & - & - & 63 days & 98 days & 102 days & - & - \\
\textbf{HIRES-post Data} \\
~~~~ Median Precision & 1.10 m s$^{-1}$ & 2.44 m s$^{-1}$ & 2.38 m s$^{-1}$ & 0.86 m s$^{-1}$ & 1.2 m s$^{-1}$ & 1.17 m s$^{-1}$ & - & - \\
~~~~ N$_{RV}$ & 23 & 19 & 33 & 97 & 4 & 68 & - & - \\
~~~~ RV baseline & 3362 days & 3379 days & 861 days & 3604 days & 2301 days & 5665 days & - & - \\
~~~~ Median Cadence & 1 day & 5 days & 2 days & 7 days & 614 days & 39 days & - & - \\
\textbf{HARPS Data} \\
~~~~ Median Precision & - & - & - & - & - & - & 2.8 m s$^{-1}$ & 0.84 m s$^{-1}$ \\
~~~~ N$_{RV}$ & - & - & - & - & - & - & 50 & 35 \\
~~~~ RV baseline & - & - & - & - & - & - & 95.8 days & 4400 days \\
~~~~ Median Cadence & - & - & - & - & - & - & 1 day & 8 days \\
\textbf{HARPS-N Data} \\
 ~~~~ Median Precision & - & 1.23 m s$^{-1}$ & 0.91 m s$^{-1}$ & - & - & - & - & - \\
 ~~~~ N$_{RV}$ & - & 77 & 104 &  &  &  & - & - \\
~~~~ RV baseline & - & 1971 days & 1974 days & - & - & - & - & - \\
~~~~ Median Cadence & - & 1 day & 1 day & - & - & - & - & - \\
\textbf{APF Data} \\
 ~~~~ Median Precision & 0.99 m s$^{-1}$ & - & - & - & - & - & - & - \\
 ~~~~ N$_{RV}$ & 61 & - & - &  &  &  & - & - \\
~~~~ RV baseline & 1902 days & - & - & - & - & - & - & - \\
~~~~ Median Cadence & 4 days & - & - & - & - & - & - & - \\
\textbf{NEID Data} \\
 ~~~~ Median Precision & 0.52 m s$^{-1}$ & 2.01 m s$^{-1}$ & 1.36 m s$^{-1}$ & 0.56 m s$^{-1}$ & 0.39 m s$^{-1}$ & 0.71 m s$^{-1}$ & 5.2 m s$^{-1}$ & 0.73 m s$^{-1}$ \\
 ~~~~ N$_{RV}$ & 27 & 20 & 17 & 23 & 24 & 22 & 12 & 10 \\
~~~~ RV baseline & 466.7 days & 423 days & 143.8 days & 139.7 days & 120 days & 113.9 days & 60.1 days & 33 days \\
~~~~ Median Cadence & 5 days & 5 days & 6 days & 4 days & 3 days & 4 days & 3 days & 3 days \\
\enddata
\end{deluxetable*}

\subsubsection{K2 Photometry}

We utilize K2 photometry for our remaining five targets, HD 24040, HD 31966, HD 106315, HD 4256, and HD 119291. The K2 mission was a successor mission to Kepler after a failure of two of its reaction wheels prevented the spacecraft from continuing its primary mode of operation \citep{howell14}. The follow-up K2 observations started on 30 May 2014 and ended on 30 October 2018, and targeted Earth's ecliptic stars, rather than those in the Kepler field. 

The K2 observing strategy was significantly different than that of Kepler, as it could not point continuously at a single region of stars. Instead it would observe a field for $\sim$83 days and move to a different ecliptic region. Consequently, most of our targets have only a single $\sim$ 83 day span of photometry taken with K2, rather than the extensive four year time span of our Kepler targets.

K2 observations were taken in short ($\sim$1 minute) and long ($\sim$30 minute) cadence mode. We used short cadence data in our analysis when available, though, as we mention in \S \ref{sec:training}, we typically bin data into regions of size 0.1 days, and so their should be no discernable difference between cadence types.

Most important when analyzing K2 data is the treatment of systematic trends in the data. Due to the failed reaction wheels, K2 observations were consistently drifting off target, requiring thruster fires to keep stars in the field of view \citep{mayo18}. This would consistently change the pixel that each star fell on, introducing myriad systematic effects into the raw data.

A variety of different correction methods were devised for the purpose of removing systematic trends. In our analysis we utilize \texttt{EVEREST} \citep{luger16} for all of our K2 targets, with the exception of HD 4256, as the default \texttt{EVEREST} reduction included with \texttt{lightkurve} was hardly improved from the raw flux. For HD 4256 we utilize pixel level decorrelation (PLD) built into the \texttt{lightkurve} software package to produce a lightcurve that saw much reduction in the artificial signals common in K2 photometry \citep{lightkurve18}. 

With the exception of HD 119291, all of our K2 targets still showed signs of long-period systematic contamination after \texttt{EVEREST}, or PLD, corrections. To remove these signals, we used a 1D spline in the \texttt{SciPy} interpolation module, \texttt{scipy.interpolate.BSpline} \citep{virtanen20}. The smoothing factor, s, is a dimensionless parameter that controls the closeness (low values) and smoothness (high values) of the spline fit. We used large smoothing factors to prevent the spline from overfitting and removing short-period signals, typically ranging from $10^7$ - $10^9$. 

It is likely that the K2 photometry, after \texttt{EVEREST} and spline corrections, still contains systematics. Nonetheless, we go forward with our analysis with these first and second-order corrections: a more in-depth reduction would be beyond the scope of our analysis.

\subsubsection{TESS Photometry}

The Transiting Exoplanet Survey Satellite \citep[TESS;][]{ricker15} began its primary mission on 18 April 2018, and continues to take data today. While the primary purpose of the TESS mission is to find transiting exoplanets, much like Kepler and K2, its observing strategy is significantly different. TESS is an all-sky survey, starting in the Southern Hemisphere and pointing at most stars for only a $\sim$ 27 day sector before moving on to another region of sky. While not as precise as Kepler, the TESS mission is especially focused on finding exoplanets orbiting the brightest, nearest stars. To date, TESS has confirmed 446 transiting planets, though over 7,000 candidates are currently under some form of study (Taken from the NASA Exoplanet Archive on 25 May 2024). 

Because of its all-sky nature, most bright stars will receive TESS observations at some point. All of our targets were chosen during NEID observing semesters where simultaneous NEID RVs and TESS photometry were available. Because of their brightness, all of our targets recieved observations at two-minute cadence, and some even at faster cadence.

We use TESS Pre-Search Data Conditioning Simple Aperture Photometry (PDCSAP) flux processed by the TESS Science Processing and Operations Center \citep[SPOC;][]{jenkins16} pipeline. This processing removes troublesome features from raw flux data such as instrumental effects and scattered light. However, it also tends to remove astrophysical signals longer than about half a TESS sector in period-space. This might plausibly hinder TESS's utility as a training dataset for our RV analysis, and is one of the key differences between TESS and Kepler data. We include a summary of TESS precision for each of our targets in Table \ref{tab:observations}.

\subsection{Radial Velocity Data}

RV data utilized consists of archival data, as well as newly acquired observations. RVs taken with the NEID spectrometer and the Automated Planet Finder (APF) spectrograph (detailed in \S \ref{sec:neid} and \S \ref{sec:apf}) are all newly utilized, and non-archival. All other instrument data is purely archival. We detail each instrument's RV data in more detail ahead.

\subsubsection{RVs with Keck/HIRES}
\label{sec:hires}

All of our targets, with the exception of HD 119291 and HD 106315, saw observations taken using the High Resolution Echelle Spectrometer \citep[HIRES;][]{vogt94} located on the Keck 1 telescope in Hawaii. Precise radial velocities are taken from \citep{butler17}.

The HIRES spectrograph recieved an instrument upgrade in 2006 that likely introduced a systematic offset between data taken before and after. In our analysis, HD 173701, HD 4256, HD 24040, and HD 31966 all have HIRES observations on either side of this maintenance. Consequently, when modeling the data, we treat pre-upgrade HIRES data and post-upgrade HIRES data as separate instruments. HD 173701's pre-upgrade HIRES data appeared to exhibit a strong trend not seen in any of the other data, which we deem likely systematic in origin. We consequently exclude it from the analysis.

Our observation statistics are detailed in Table \ref{tab:observations}.

\subsubsection{RVs with HARPS-N}

Kepler-21 and Kepler-37 were both observed using the High-Accuracy Radial velocity Planet Searcher-North \citep[HARPS-N;][]{cosentino12} spectrograph. HARPS-N is located at the Telescopio Nazionale Galileo, a 3.6-m telescope in the Canary Islands, Spain. 

\cite{rajpaul21} utilized a pairwise Gaussian Process reduction during their analysis of Kepler-37, producing a different set of HARPS-N RVs. Such a process was not used in \cite{bonomo23}, though they include additional HARPS-N RVs. We utilize the data from \cite{bonomo23} for all of our targets. In particular, we do \textit{not} uses Kepler-37 data from \cite{rajpaul21} for 1) consistency with our other targets, and 2) we are interested in studying post-processing methods for removing stellar variability, while \cite{rajpaul21} utilized a pre-processing method to do so.

Our observation statistics are detailed in Table \ref{tab:observations}.

\subsubsection{RVs with HARPS}

Two of our targets, HD 106315 and HD 119291 were observed by the High Accuracy Radial velocity Planet Searcher \citep[HARPS;][]{mayor03}. HARPS is a high precision spectrograph utilized by the European Southern Observatory 3.6 m telescope, located at La Silla Observatory, Chile. HARPS is capable of achieving a spectral resolving power of $R \sim 115,000$. 

Our observations statistics are detailed in Table \ref{tab:observations}.

\subsubsection{RVs with the Automated Planet Finder}
\label{sec:apf}

We obtained RVs of HD 173701 using the Levy Spectrometer on the 2.4\,m Automated Planet Finder \citep[APF;][]{vogt2014} Telescope. APF is a fully robotic telescope at Lick Observatory on Mt.~Hamilton, CA.

Our APF observations of HD 173701 are comprised of 61 observations taken over a 1902-day baseline extending from July 2013 to September 2018.  We used an exposure meter to achieve consistent SNR sufficient to achieve $\sim 1$ m s$^-1$ precision; exposure times range from 322-1800s.  

The Levy spectrometer is equipped with an iodine vapor cell to provide a stable wavelength reference and track variations in the instrument profile.  RVs were extracted using the method described in \citet{butler17}.  The APF RV time series has an RMS scatter of 4.7 m s$^-1$ and a median uncertainty of 0.99 m s$^-1$.

\subsubsection{RVs with the NEID Spectrometer}
\label{sec:neid}

We obtained RV observations for all targets with the extremely precise NEID spectrograph, located on the WIYN 3.5 m telescope at Kitt Peak National Observatory\footnote{The WIYN Observatory is a joint facility of the University of Wisconsin–Madison, Indiana University, NSF NOIRLab, the Pennsylvania State University, Purdue University, and Princeton University.} \citep{schwab16} . NEID is extremely stable, with a resolving power of $R >100\,000$ and capable of achieving RV precisions as low as 25 cm s$^{-1}$. RVs were obtained via the standard NEID reduction pipeline, which utilizes a cross correlation function \citep[CCF;][]{angladaescude12} to generate precise RVs.

One of the main goals of our analysis was to explore how photometry can help mitigate stellar activity in the era of extreme precision radial velocities (EPRV). Thus, obtaining NEID data simultaneously with TESS was an essential part of our experimental design.

Obtaining RV observations simultaneously with any photometric instrument can be very challenging, but is especially so for TESS. The TESS observing strategy typically only observes a single star for a 27 day sector, limiting the ability to obtain RVs simultaneously. This is complicated by the fact the NEID instrument is not used every night at the WIYN telescope, and that scheduled NEID observations are often competing with other programs for limited observing time. The difficulty is alleviated by the queue-observing system employed by WIYN staff, allowing for high-cadence observations that are impossible in a classically scheduled system \citep{golub20}. 

Our observing programs faced several challenges. We were originally allocated NEID time during Sector 26 of the TESS primary mission, the first sector where Kepler stars were to be observed and NEID was operational. However, Kitt Peak underwent a full observatory shut down due to the SARS Cov-2 pandemic during Sector 26, making data acquisition impossible. Observations for our program continued nominally in 2021, thanks in part to the NEID queue system, and we initially had great success observing during Sectors 40 and 41. The latter saw a particularly bad patch of weather at Kitt Peak, and few of our final observations were able to be executed. 

\begin{figure*}
    \centering
    \includegraphics[width=\textwidth]{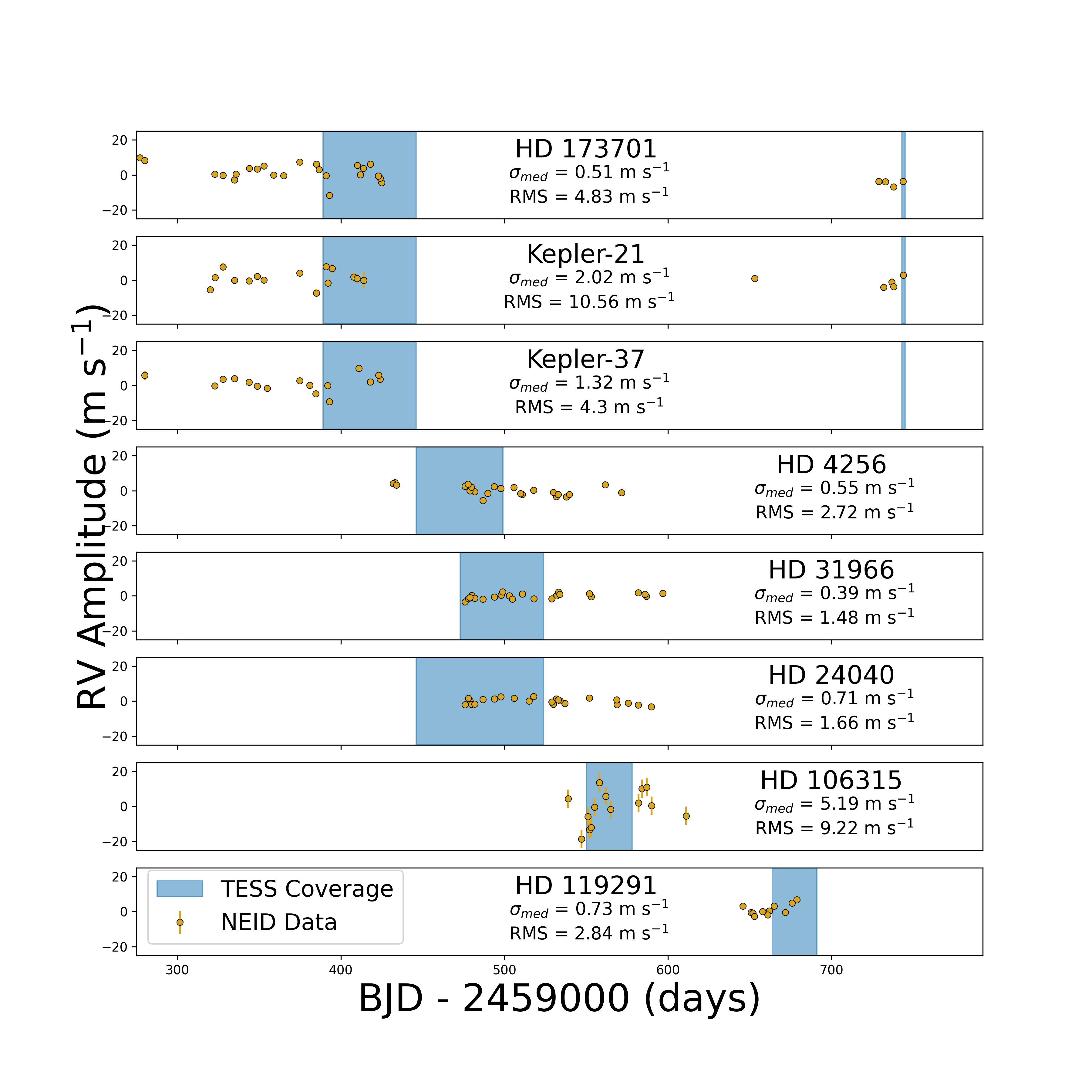}
    \caption{NEID observations of our eight targets. TESS observations are overlaid in blue. Bad weather hindered observations of Kepler targets at first, and the Contreras fire would eventually prevent observations toward the end of our program, hence the small window of TESS observations which then ended abruptly for the Kepler targets. We generally had high success observing K2 targets.}
    \label{fig:neid}
\end{figure*}

TESS then moved to the Southern hemisphere for year three of its mission (Sectors 27-39), and we could no longer observe Kepler targets simultaneously with TESS. Fortunately, some ecliptic targets could be observed with NEID and TESS simultaneously, and we obtained RVs of our K2 targets. After a year in the Southern Hemisphere, TESS would return to the North and the Kepler field. As is visible in Figure \ref{fig:neid}, these observations quickly ceased due to the Contreras wildfire, which threatened Kitt Peak National Observatory, halting observations for several months.

Despite the many setbacks our program faced, we obtained many NEID RVs simultaneously, or nearly simultaneously, with TESS observations. We include a plot of our simultaneous observations in Figure \ref{fig:neid}.

Our observation statistics are detailed in Table \ref{tab:observations}.

\section{Analysis}\label{sec:analysis}

\subsection{Training Activity Models}
\label{sec:training}

Much of our analysis is devoted to determining which photometric datasets can be used to glean the most information about stellar activity contamination in the RVs. Kepler data has a much longer baseline than TESS, but is comparatively old, and may not well predict the current stellar activity structure. K2 data, while longer baseline than TESS, has its own disadvantages, as the data include difficult-to-remove instrumental signals that might negate any advantage from the longer observing baseline. Future photometric missions such as PLATO \citep{rauer14} will likely provide different advantages and disadvantages still. To the end of quantifying advantages and disadvantages, we carry out analyses that are trained on Kepler/K2 photometry, or TESS photometry, as well as untrained analyses, and scrutinize the results.

Our training involves fitting an activity model to each photometric dataset, and using Markov-chain Monte Carlo \citep[MCMC;][]{foreman-mackey13} sampling to determine the best fit posteriors. We use these posteriors as priors when performing RV-only fits, or when performing injection-recovery fits. This approach is one method for utilizing photometry to inform stellar activity models, though it is not the only method, or necessarily the best \citep{nicholson22,tran23}. Nonetheless, the method is widely used and generally reliable. We choose as our activity model the $\mathcal{K}_{J1}$ chromatic GP kernel, detailed in \cite{cale21}. This kernel is an expansion of the commonly used quasi-periodic (QP) kernel, with a few advantages not implemented in the QP kernel by default. The $\mathcal{K}_{J1}$ kernel is given in equation \ref{eqn:kj1}.

\begin{equation}
    \label{eqn:kj1}
    \centering
    \scriptsize
    \mathcal{K}_{J1} = \eta_{1,s(i)}\eta_{1,s(j)}\exp \bigg(\frac{-|t_{i} - t_{j}|^{2}}{2\eta^{2}_{2}} - \frac{1}{2\eta^{2}_{4}}\sin^{2}\big(\frac{\pi|t_{i}-t_{j}|}{\eta_{3}}\big)\bigg)
\end{equation}

The first advantage we expect from this kernel is that it utilizes a different activity amplitude term for each instrument, which is generally physically motivated, as different RV and photometric instruments often extract information from different bandpasses, and stellar activity is chromatic \citep{crockett12}. Additionally, the $\mathcal{K}_{J1}$ GP kernel models all instruments in a single covariance matrix, allowing covariances \textit{between instruments} to be enforced, whereas the traditional QP kernel models each instrument independently.

The advantages of the $\mathcal{K}_{J1}$ kernel are not as obvious when training on photometric data, but we expect real benefit to come from RV fits utilizing this kernel. For consistency, we do all our training and model evaluation using this kernel, though on single-instrument photometry, the $\mathcal{K}_{J1}$ kernal is identical to the QP kernel. The calculation of the chromatic GP likelihood scales with number of data points by $\mathcal{O}(${N$^{3}$}), making the computation impractical for very large datasets, like photometry. To circumvent this problem, we train on binned subsets of photometry. We use 30 minute time bins, as RV cadence is much sparser than this, and stellar activity that evolves on shorter timescales is not likely to be detected in the RVs. 

We additionally use subsets of the photometric data to speed computation. For Kepler, we chose quarters 6 and 7 for this analysis. This choice was originally motivated by the availability of simultaneous Kepler-21 RV data in these quarters, and we continue to adopt it for our other Kepler targets for the sake of consistency. K2 targets saw much shorter baseline observations, and we use all available K2 data for each target.

TESS photometry for some targets is highly abundant, but often spread out over many years with large gaps of no observation. During survey development, we chose targets where simultaneous NEID observations could be acquired with TESS observations. Consequently, we generally choose TESS photometry closest to our NEID observations for photometric training. A full table of which datasets for which targets were used for training is given in Table \ref{tab:training}.

\begin{deluxetable}{cccc}
 \label{tab:training}
\tablecaption{Datasets Used for Photometric Training}
\tablehead{\colhead{Target}  &  \colhead{Kepler}
& \colhead{K2} & \colhead{TESS}}
\startdata
HD 173701 & Q6, Q7 & - & S40, S41 \\
Kepler-21 & Q6, Q7 & - & S40, S41 \\
Kepler-37 & Q6, Q7 & - & S40, S41 \\
HD 4256 & - & C8 & S42, S43 \\
HD 31966 & - & C13 & S43, S44 \\
HD 24040 & - & C4 & S42, S43, S44 \\
HD 106315 & - & C10 & S46 \\
HD 119291 & - & C17 & S50 \\
\enddata
\tablenotetext{}{Q is short for Kepler quarter, C for K2 observing campaign, and S for TESS sector.}
\end{deluxetable}

We chose broad, uninformative priors for our fits, to maximize the amount of ``learning" that could be done. For $\eta_{1}$, we implemented a Jeffreys prior with minimum 10 ppm, and maximum 1,000,000 ppm. For $\eta_{2}$, we implemented a wide Jeffreys prior with a minimum equal to the predicted rotation period, and a maximum of 10,000 days, as suggested by Polanski et al., 2024 (submitted). For $\eta_{3}$, which is often a good approximation of the stellar rotation period, we utilize a Gaussian prior centered at the predicted rotation period (Table \ref{tab:stellar}), and with a width of 20$\%$ of this estimate. Finally, we follow \cite{LM16} and Polanski et al., (submitted) concerning $\eta_{4}$ by utilizing a Gaussian prior for centered at 0.5, with a standard deviation of 0.05.

We run each training dataset through an MCMC inference process to measure, primarily, its GP hyperparameters. We generate an activity-only GP model using the \texttt{RadVel} software package \citep{fulton18}. We follow the default \texttt{RadVel} convergence criteria to assess convergence, which assesses convergence by determining when the Gelman-Rubin (G-R) statistic \citep{ford06} is less than 1.01 and the number of independent samples is greater than 1000 for all free parameters for at least five consecutive checks. After inference is completed, we use the posteriors of all the hyperparameters as priors for our RV fits, with the exception of amplitude. This is because photometry and RV data have completely different dimensions, and photometric amplitude cannot be reliably converted into RV amplitude. We implement posteriors as priors by taking the posterior mean and standard deviation, and using these as Gaussian priors in our RV fits. A summary of our training posteriors is given in Table \ref{tab:train_vals}.

\begin{deluxetable}{cccc}
\label{tab:train_vals}
\tablecaption{Trained Values}
\tablehead{\colhead{Target}  &  \colhead{Kepler}
& \colhead{K2} & \colhead{TESS}}
\startdata
\textbf{HD 173701} &  &  &  \\
~~~$\eta_{2}$ (days) & 41.7$\pm$0.5 & - & 51$\pm$11 \\
~~~$\eta_{3}$ (days) & 32.81$\pm$0.07 & - & 42.0$\pm$5.6 \\
~~~$\eta_{4}$ & 0.098$\pm$0.003 & - & 0.049$\pm$0.009 \\
\textbf{Kepler-21} &  & &  \\
~~~$\eta_{2}$ (days) & 17.5$\pm$1.6 & - & 15.7$\pm$1.3 \\
~~~$\eta_{3}$ (days) & 22.03$\pm$0.11 & - & 15.8$\pm$2.3 \\
~~~$\eta_{4}$ & 0.090$\pm$0.003 & - & 0.14$\pm$0.05 \\
\textbf{Kepler-37} &  & &  \\
~~~$\eta_{2}$ (days) & 27.7$\pm$1.2 & - & 30$\pm$4 \\
~~~$\eta_{3}$ (days) & 26.5$\pm$0.2 & - & 17.6$\pm$0.22 \\
~~~$\eta_{4}$ & 0.22$\pm$0.01 & - & 0.27$\pm$0.07 \\
\textbf{HD 4256} & & & \\
~~~$\eta_{2}$ (days) & - & 55$\pm$1 & 60$\pm$15 \\
~~~$\eta_{3}$ (days) & - & 50$\pm$6 & 18.4$\pm$0.3 \\
~~~$\eta_{4}$ & - & 0.5$\pm$0.05 & 0.42$\pm$0.05 \\
\textbf{HD 31966} & & & \\
~~~$\eta_{2}$ (days) & - &  31.9$\pm$3.2 & 809$^{+2100}_{-809}$ \\
~~~$\eta_{3}$ (days) & - & 26.0$\pm$0.6  & 27.9$\pm$5.4 \\
~~~$\eta_{4}$ & - & 0.42$\pm$0.05 & 0.51$\pm$0.05  \\
\textbf{HD 24040} & & & \\
~~~$\eta_{2}$ (days) & - & 38$\pm$12 & 66$\pm$29 \\
~~~$\eta_{3}$ (days) & - & 29.8$\pm$2.8 & 8.35$\pm$0.09 \\
~~~$\eta_{4}$ & - & 0.52$\pm$0.05 & 0.48$\pm$0.05 \\
\textbf{HD 106315} & & & \\
~~~$\eta_{2}$ (days) & - & 6.2$\pm$0.37 & 9$^{+51}_{-9}$ \\
~~~$\eta_{3}$ (days) & - & 5.36$\pm$0.64 & 4.52$\pm$0.41 \\
~~~$\eta_{4}$ & - & 0.51$\pm$0.05 & 0.48$\pm$0.05 \\
\textbf{HD 119291} & & & \\
~~~$\eta_{2}$ (days) & - & 41.9$\pm$2.7 & 61$\pm$24  \\
~~~$\eta_{3}$ (days) & - & 24.71$\pm$0.15  & 14.86$\pm$0.15 \\
~~~$\eta_{4}$ & - & 0.26$\pm$0.02 & 0.28$\pm$0.15 \\
\enddata
\caption{$\eta_{2}$ approximates the spot-decay lifetime on the surface of a star, $\eta_{3}$ is an approximation of the stellar rotation period, and $\eta_{4}$ is the periodic scale length, which roughly controls the intra-period variations of the stellar activity.}
\end{deluxetable}

\subsection{Injection-Recovery Tests}

We utilize an injection-recovery test to explore which training method best removes stellar activity contamination. For each target, we take its RV data and inject a variety of planetary signals with known orbital period, RV semi amplitude, and phase. We then fit an RV model with the injected planet included and an RV model with it excluded, and compare the results. For systems with known planets, the injected signal is fit in addition to the known planet or planets, with the exception of Kepler-37, where only the largest of the known planets is included, as the others are below our sensitivity. We then elect to use model comparison to determine with what level of confidence that the injected planet is recovered. 

Another possible metric for model effectiveness is the recovered RV semi-amplitude of a simulated planet, divided by its uncertainty. However, due to the high computational cost of our injection-recovery tests detailed later in the section, we were unable to perform true inference, and thus could not achieve a reliable estimate of our amplitude uncertainty. Consequently, we opt to use model comparison rather than recovered amplitude significance. 

We follow a similar method as used in \cite{cale21} by injecting planets with a variety of periods and amplitudes. We create 10 bins linearly spaced between 1 and 10 m s$^{-1}$ for RV amplitudes, and 10 bins with log-uniform spacing between 0.1 days and 1000 days for injected orbital period. The result is 100 bins for a range of amplitudes and orbital periods. When injecting signals, for each bin, we generate a fake planet with RV amplitude and orbital period randomly selected inside each bin, and with zero eccentricity.  Additionally, we then randomly pick a phase for the injected planet to prevent our results from being biased by RV phase coverage. We do this 100 times in each bin, and average the results.

Following \cite{cale21}, we explore two cases: first, where the injected planet is ``transiting," and we know the orbital period and time of conjunction precisely. In these cases, these parameters are fixed when fitting, and only the RV amplitude $K$ of the injected planet is fit. The second case is an ``RV detected" planet, where we do not know the precise orbital period or phase, and these parameters must also be fit.

To determine the effectiveness of our training, we compare the results of RV fits that include the injected planet, and those that do not. To do so, we perform N and N+1 planet fits for each system and training dataset, where N is the number of planets known in the system. We compare the results of the fit using the evidence of each model, which is calculated by integrating the product of the model likelihood and its priors over the entire parameter space \citep{kass95}. A higher evidence value indicates a better fit to the data, and we use the concept of the Bayes Factor (BF) to quantify the improvement gained by adopting the higher evidence model. This is usually defined as the ratio of the evidences, though it is more often the logarithm of the BF that is computed and referenced, as the log of the evidence is much less likely to overflow a computer's floating point precision.

The primary issue with utilizing evidences to explicitly calculate the BF is that they are very computationally expensive, and often completely impractical to estimate. A variety of methods exist for approximating the evidence or the BF \citep[BIC, AIC, Nested Sampling;][]{liddle07,espinoza19}, though all have their drawbacks. BIC and AIC typically require a full MCMC sampling process before they can be reliably calculated, and even then are considered imperfect approximations of the evidence. Nested sampling can be used to calculate the evidence of a model without taking the complicated integral, but is itself a sampling process that can be time-consuming.

The problem is especially severe in our case, as our injection-recovery tests are performed for 100 different grids, and with 100 simulated phases inside each grid. These 10,000 calculations have to be performed six times for each target: fits without training, trained on Kepler/K2, trained on TESS, and in each case fits including the injected planet in the model, or not. Thus, 60,000 model evidences have to be calculated for each of our targets, a totally infeasible task using the above methods.

Instead, we approximated the evidence using the Laplace approximation (detailed in \cite{nelson20} A3). Essentially, the Laplace approximation leverages the fact that a complicated integral, decomposed into an exponential

\begin{equation}
    Z = \int \exp(f(x)) \, dx
\end{equation}

can be approximated as

\begin{equation}
    \bigg[\frac{(2\pi)^{2}}{\det|H(x_{0})|}\bigg]^{\frac{1}{2}} \times \exp(f(x_{0}))
\end{equation}

where $H$ is the Hessian matrix of $f(x)$, and $x_{0}$ is the dominant posterior mode. This approximation is only valid if this dominant posterior mode is far from the bounds of integration, and if the likelihood is concentrated unimodally around $x_{0}$. We performed a series of tests to verify the validity of this approximation on the systems we analyze. First, we compared, for each fit type, the recovered planetary amplitudes to those that were injected, and verified that they were generally within 20$\%$ of the true values. This indicates that our fits were optimizing model parameters to the dominant posterior mode. Second, we performed full MCMC fits on a small subset of our injected datasets to verify that posteriors were unimodal around this region. The full details of our checks are available upon request. We utilize a lightweight python package, \texttt{LApprox} \citep{beard24_lapprox}, to estimate the evidence for each of our runs.

It is important to caution the reader's interpretation where injected amplitudes are low, and injected periods are long. At typical RV cadences and precisions, such signals are challenging to recover using any method, and are more susceptible to incorrect fitting and posterior estimations. Additionally, the Laplace approximation may be less valid in these regions.

\begin{figure*}
    \centering
    \includegraphics[width=\textwidth]{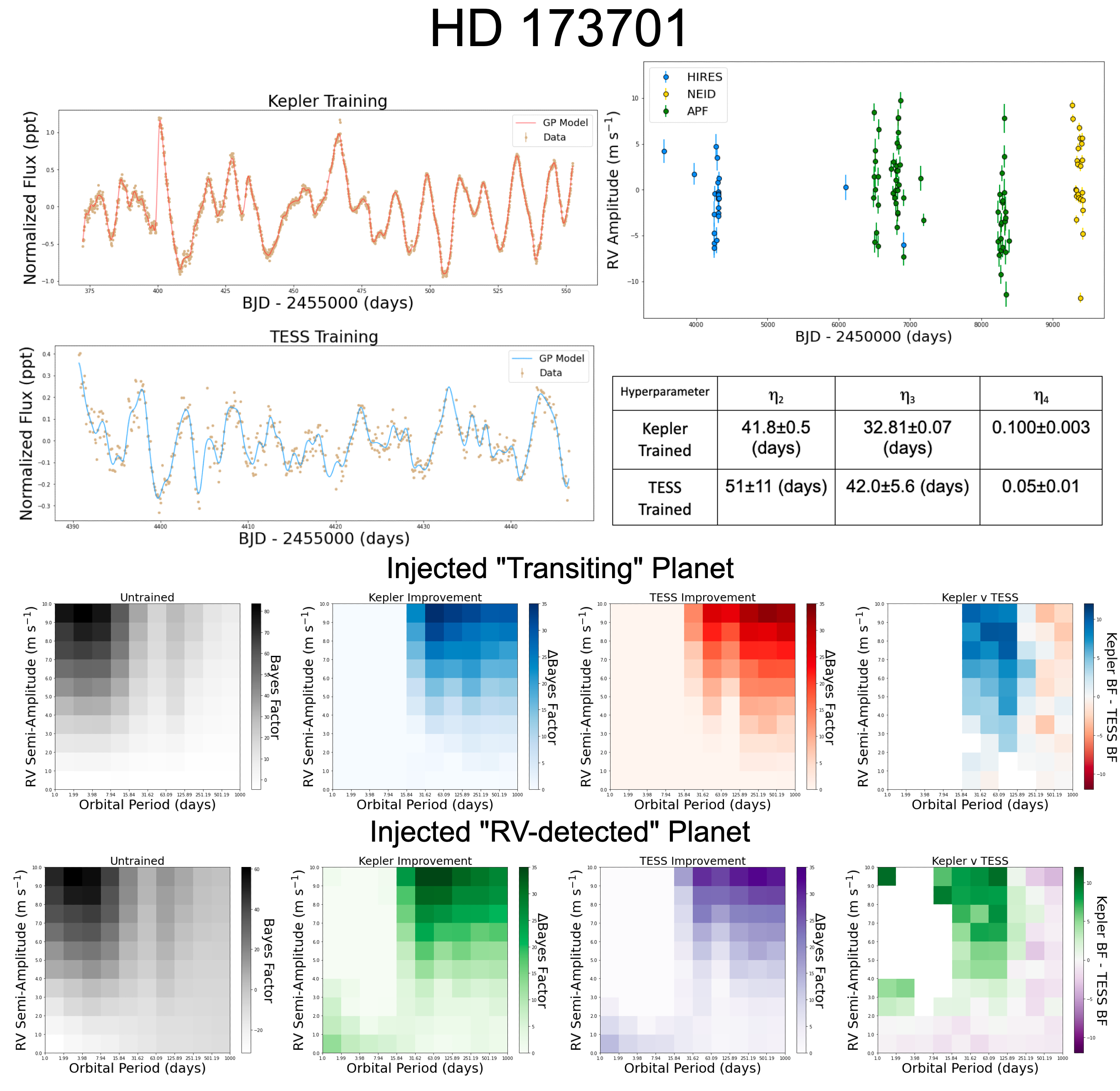}
    \caption{We include a variety of plots summarizing our training and analysis of HD 173701. Top Left: Kepler and TESS training data, as well as our best fit GP model overlaid. Top Right: RV time series and training posteriors. Bottom: Results of our injection-recovery analysis in the two cases described in \S \ref{sec:analysis}. The left plots show the preference for models including the injected planet when no GP training is applied. The middle two plots show the improvements gained when training on Kepler or TESS. The rightmost plots highlight the differences between Kepler and TESS training.}
    \label{fig:HD173701_card}
\end{figure*}

We include a summary figure for each of our targets in the appendix, though highlight HD 173701 in the main text in Figure \ref{fig:HD173701_card}. These include plots of our GP fits to the photometry, as well as summary plots of the injection recovery results for the ``transiting" and ``RV-detected" cases. 

\subsection{Non-Simultaneous TESS Fits}
\label{sec:nonsim}

Beyond determining which training dataset improves RV fits most significantly, we are interested in quantifying the effectiveness of simultaneously obtained RV data. Such data is widely considered ideal for constraining RV activity models using photometry, but is often difficult to acquire in practice. 

To do so, we perform another injection-recovery test on a model that has been trained on non-simultaneous TESS photometry. In Table \ref{tab:training}, we note the TESS sectors that were used for training our activity models in our main analysis. These were selected because they were the TESS sectors simultaneous with our acquired NEID RV data, and we hypothesized that a model trained on this data would be most effective at separating stellar activity signals from exoplanet signals. To test this hypothesis, we train on non-simultaneous TESS sectors, and we compare the result. HD 106315 and HD 119291 have no additional TESS photometry, and could not be included in this analysis, but we were still able to analyze six targets. We used sectors 53 and 54 for the three Kepler targets, sector 70 for HD 4256, sector 71 for HD 31966, and sectors 70 and 71 for HD 24040. For the Kepler targets, this is a median RV-photometry separation of 356, 352, and 362 days for HD 173701, Kepler-21, and Kepler-37, respectively. For the K2 targets, this corresponds to a median RV-photometry distance of 709 days for HD 4256, 719 days for HD 31966, and 684 days for HD 24040.

\begin{figure*}
    \centering
    \includegraphics[width=\textwidth]{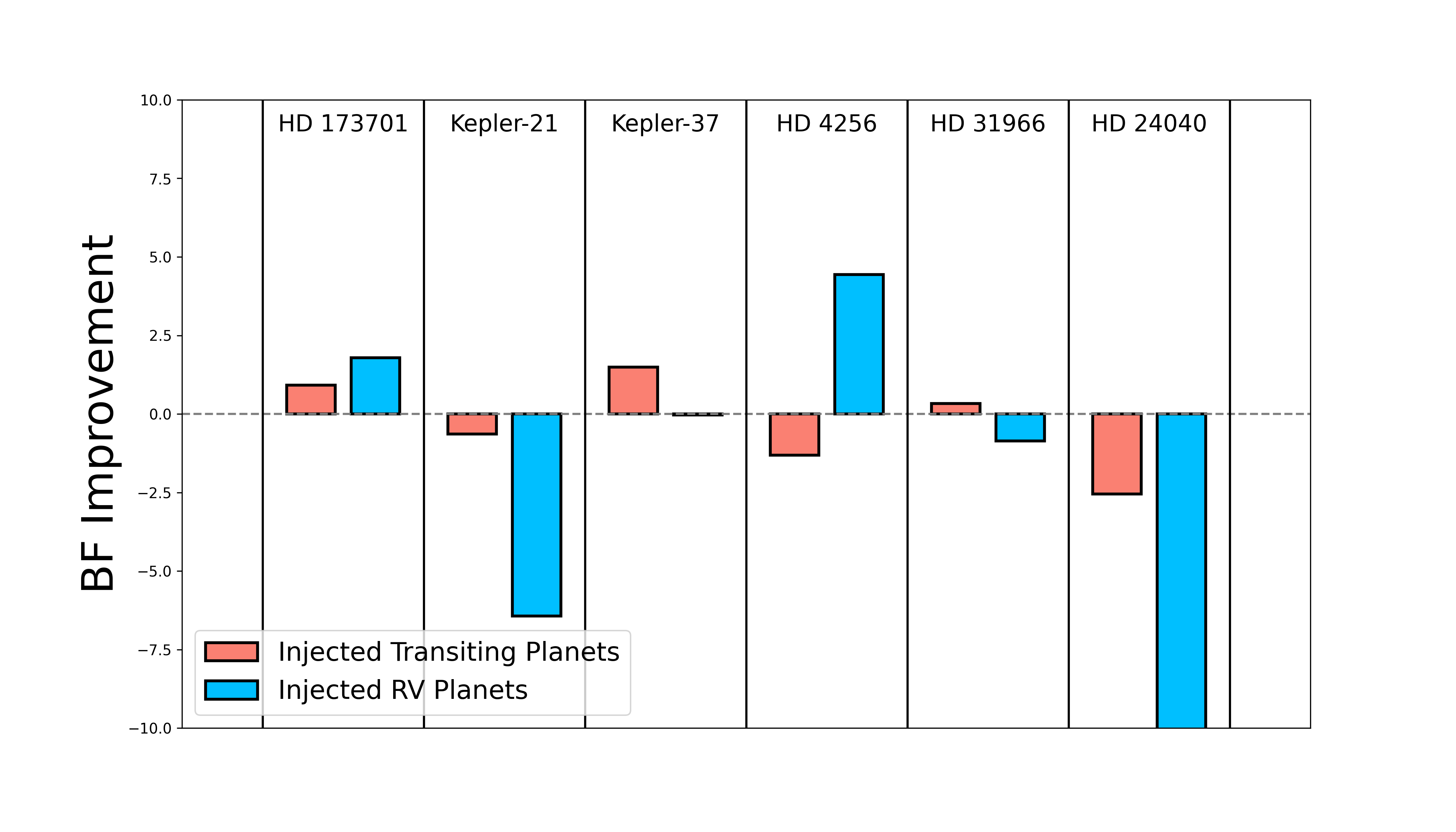}
    \caption{We show the increased or decreased performance of our activity models when trained on simultaneous data instead of non-simultaneous data. The y-axis indicates $\Delta$BF between models trained on simultaneous versus non-simultaneous photometry. Positive values indicate that the simultaneous photometry is improving sensitivity to injected planets, while negative values indicate worse performance. Red bars correspond to our ``transiting" planet runs, and the blue bars correspond to ``RV-detected" injected planets. There is not consistent improvement gained from simultaneous photometric training. We note that for HD 24040, the negative preference goes far below our axes limits, which we set for a clearer analysis of the other systems. As mentioned in the text, we believe these fits may not be reliable.}
    \label{fig:simultaneity_improvement}
\end{figure*}

We train activity models on these non-simultaneous TESS sectors just as in \S \ref{sec:training}, and we compare the differences in effectiveness in Figure \ref{fig:simultaneity_improvement}.

\section{Discussion}
\label{sec:discussion}

\subsection{Which Photometric Dataset?}

Our injection-recovery tests shine some light on the advantages of training on a photometric dataset. Furthermore, we highlight the strengths of either Kepler or TESS training. We provide summary plots for each of our eight targets in Figures \ref{fig:HD173701_card} and \ref{fig:Kepler-21} through \ref{fig:HD119291}. These plots contain the Kepler and TESS training timeseries, the RV time series, the results of our training, and the results of injection-recovery tests. We focus on the bottom two panels of each plot in this section.

The leftmost injection-recovery panels show the level of preference for a fit with the injected planet over a fit without the injected planet when the GP model is not trained. We use the Bayes Factor (BF) as described in \S \ref{sec:analysis} to quantify this value. The middle panels quantify the level of improvement gained by training on Kepler or on TESS, as compared to no GP training at all. Finally, the rightmost panels show the difference between the middle two, emphasizing which photometric dataset improves the BF by the greatest amount. We discuss each target briefly before our final summary just after.

\subsubsection{HD 173701}

The results of our HD 173701 analysis are visible in Figure \ref{fig:HD173701_card}. The system is one of our brightest targets, and its photometry exhibits very clear quasi-periodic modulation. We predict a rotation period of 41.38 days from the $\log R^{\prime}_{HK}$ value in \S \ref{sec:training}, though GLS and autocorrelation analysis of the Kepler photometry suggest a $\sim$ 32-36 day rotation period \citep{zechmeister18,holcomb22}. Like the other Kepler targets, the injected planet sensitivity is generally highest when the model has been trained on Kepler photometry, though at longer periods TESS occasionally performs modestly better. Interestingly, shorter period injected planets, especially with larger amplitudes, do not appear to benefit from training an activity model. This trend appears for several targets, and we discuss it in \S \ref{sec:all_targ}.

\subsubsection{Kepler-21}

A more detailed discussion relating to Kepler-21 is available in Beard et al. 2024b in prep, though we summarize here. The results of our analysis are visible in Figure \ref{fig:Kepler-21}. Kepler-21 exhibits a fairly clear $\sim$ 12 day rotational modulation in its Kepler photometry, though the rotation term of our GP model does not adhere to this value during training. TESS training recovers a value closer to the true rotation period, though with a larger uncertainty. Training on Kepler photometry still recovers injected planets more often than when training on TESS photometry. This suggests that some genuine stellar activity contamination likely exists near 22 days, and that this contamination exists in the RVs as well. The source of such a periodic signal is difficult to ascertain considering the system's known rotation period, though this affirms the importance of utilizing an entire dataset to mitigate stellar variability, rather than a single number for a stellar rotation period.

\subsubsection{Kepler-37}

Kepler-37 results are visible in Figure \ref{fig:Kepler-37}. The system exhibits clear rotational modulation on a $\sim$ 29 day time scale, too long for TESS to plausibly recover. As with the other Kepler targets, GP models trained on Kepler photometry consistently recover injected planets with higher confidence. 

Unlike with the other Kepler targets, training on Kepler much more significantly recovers the higher amplitude, shorter period planets injected during our analysis than when the GP model is not trained, or when it is trained on TESS. It may be that some higher frequency signal in the RVs confuses any attempt to recover planets, but is more easily distinguished when trained on Kepler photometry.

\subsubsection{HD 4256}

The K2 photometry of HD 4256 likely still contain systematic contamination, despite the efforts in \S \ref{sec:data}. The system also maintains higher levels of stellar variability in photometry and RVs, evident from TESS and S$_{HK}$ values. We expect that this contributes to the fact that TESS training seems to typically be the best at recovering unknown planets during our tests.

The system is known from its S$_{HK}$ values to have long-period activity cycles. Neither K2 nor TESS are likely to be sensitive to such long period signals, and this may explain the fairly precipitous dropoff of its sensitivity to injected planets between 60 and 100 days.

\subsubsection{HD 31966}

Our analysis of HD 31966, visible in Figure \ref{fig:HD31966}, indicates that K2 photometry is more informative for removing stellar variability from RV data. It may be that the higher precision of K2 is contributing most, as HD 31966 is one of our least active targets. TESS photometry does not show clear, quasi-periodic signals, and the amplitude of our model is near the precision TESS achieves on HD 31966. 

Strangely, the ``RV detected" and ``transiting planet" cases tell opposite stories for this system. In the former, K2 performs better when recovering injected planets between 30 and 60 days, and worse elsewhere. On the other hand, the latter sees TESS recovery improve in this region, and perform worse elsewhere. It may be that the choice of training dataset should be motivated in part by the nature of the exoplanets in the system.

\subsubsection{HD 24040}

Results pertaining to HD 24040 are visible in Figure \ref{fig:HD24040}. The target is unique among our study in that it has known, RV-detected exoplanets that do not transit. These planets are long-period, and a trend in the data suggests the possible existence of another yet. Its K2 photometry either exhibits small fluctuations due to stellar variability, or there is some residual systematic signal. TESS data, too, exhibit low levels of variability. For these reasons, the injection recovery tests are somewhat ambiguous. Kepler and TESS see preference in the ``fixed" planet case in different regimes, and rarely improve the model at all in the ``RV-detected" planet case. In fact, in the latter case, our injection recovery tests failed even when the injected signal was large. This is unexpected, and difficult to explain. Most likely, the existence of long period giant planets interferes with our ability to recover injected signals, and the issue is amplified when the period of the signal is allowed to vary.

\subsubsection{HD 106315}

HD 106315 results can be seen in Figure \ref{fig:HD106315}. Clear quasi-periodic variability in both K2 and TESS photometry can be seen, and the RV data too appear to vary with high amplitude. HD 106315 is a strange case, as it seems to have genuinely high variability in both photometry and RVs, and yet it fares better when trained on K2, rather than TESS as we might expect. The most likely culprit is the poor precision of our NEID RVs, especially as compared to our other targets. The archival HARPS data is much more precise than our NEID data, and closer in time to K2 photometry. This is likely the reason that K2 training performs better.

\subsubsection{HD 119291}

Interestingly, our analysis of HD 119291 (Figure \ref{fig:HD119291}) suggests TESS training is superior for this target. TESS photometry is not abundant, and the K2 photometry for this target appears to be less contaminated with sytematics than for some of our other targets, all which would suggest the opposite result. It seems that the simultaneous, highly precise NEID RVs contribute enough to the injection-recovery tests that TESS training is more important for this system.

It is also interesting to note a dramatic improvement in sensitivity to injected planets between 30 and 60 days, that then falls. The predicted rotation period falls in this regime, though our GP models all adhere to lower periods. Most likely, a real activity signal at this period constructively interferes with our injected planets, amplifying our sensitivity.

\subsubsection{All Targets}
\label{sec:all_targ}

First we examine the three Kepler targets, HD 173701, Kepler-21, and Kepler-37. The first conclusion that we can make is that for low-amplitude injected signals, the GP training has little effect. This is not surprising, as low-amplitude planets are challenging to recover in even a quiet dataset. In such fits, the activity model is not the dominant source of uncertainty, rather it is the number and precision of the RVs. Perhaps more surprising, all three datasets see little-to-no improvement over the untrained fits for short period, high amplitude signals (with the exception of high amplitudes for Kepler-37). Why might this be? Such injected signals are expected to be recovered most easily, and so it may be that no matter how well the activity signals is constrained, the injected planet is recovered about as well in all fits. Interestingly, for longer period injected planets, Kepler seems to be the best photometric dataset to train on, despite its age. The longer observation baseline, higher precision, and sensitivity to longer-period signals beats out any advantage recent TESS photometry might provide, despite our recent, simultaneous NEID RVs. 

The K2-trained targets paint a less clear picture. Of the five, HD 4256 and HD 119291 have a clear preference for TESS training, while HD 31966 and HD 106315 benefit more from K2 training. HD 24040 has no clearly winning dataset. HD 106315's preference may be the easiest to explain: its NEID RVs are the least precise in our dataset, and it seems plausible that the advantages of training on TESS (mainly simultaneity with NEID) are diminished when the NEID data are less constraining of the RV model. HD 119291 and HD 4256 both have highly precise NEID RVs, likely emphasizing the importance of NEID simultaneity. HD 31966 is confusing from this angle, however, as its NEID RVs are the most precise in the dataset, and we have achieved the largest number of simultaneous RVs. Perhaps the best explanation is the fact that it has the smallest amplitude of activity contamination, similar to HD 24040. Examining the $\log R^{\prime}_{HK}$ values in Table \ref{tab:stellar}, this seems probable. $\log R^{\prime}_{HK}$ values can be a good measure of stellar activity \citep{dasilva11}, and HD 119291 and HD 4256 have the largest values in the K2 target list. The advantages of training on simultaneous photometry seem to diminish as activity contamination becomes smaller.

Comparing the ``transiting" planet fits to the ``RV detected," we can see that GP training has a larger effect, typically, for the latter case across the board. This is unsurprising: when the period and phase of the undetected planet is uncertain, it is more difficult for an untrained GP to distinguish stellar activity from the injected planet. Thus, training a GP on photometry has extra benefit when searching for non-transiting planets.

Kepler and TESS photometry have many different qualities. Kepler photometry is much longer in baseline, higher in precision, and sensitive to longer period stellar activity signals, but is temporally distant from many of our most precise RVs. Despite this, training an activity model on Kepler photometry consistently recovers injected planet signals more strongly than when activity models are trained on TESS. This sensitivity increase seems to uptick above half of a TESS sector length, which may suggest that the short baseline of TESS observations is the largest hindrance to mitigating activity. 

K2 targets often better recover injected planets when trained on TESS. The K2 mission has significantly shorter baseline than Kepler (typically $\sim$ 80 days), though it is still much longer than a typical TESS sector. K2 data are also notoriously contaminated with systematic effects. A GP activity model, when trained on such a dataset, might attempt to remove signals that originate from K2 systematic effects from RV data, despite their absence. This, combined with K2's shorter baseline when compared to Kepler, seems to hinder K2's other advantages.

\subsection{Are Simultaneous RVs Beneficial?}
\label{sec:beneficial}

We are also interested in exploring the benefit of simultaneously acquired RVs. It is commonly expected that RVs acquired simultaneously with photometry are ideal, as photometry can be used to constrain stellar activity signals in RVs as they are happening. Simultaneous RVs can be very challenging to acquire, however, and their true benefit may be small. Many analyses today utilize disparate datasets with a variety of RV and photometric instruments, further diminishing the probable benefit of simultaneously acquired RVs.

We performed a series of injection-recovery tests trained on simultaneous TESS photometry, as well as on non-simultaneous TESS photometry (refer to \S\ref{sec:nonsim} for details). We take the median BF preferring the injected-planet model from our simultaneous-TESS and non-simultaneous-TESS analyses, and we use these numbers as a metric of how well the training is identifying injected planets across period-$K$ amplitude space. We then take the difference between the simultaneous and non-simultaneous BFs in Figure \ref{fig:simultaneity_improvement}. 

A clear pattern is hard to distinguish. Simultaneous photometry is \textit{not} the clearly better dataset with which to train an activity model. HD 173701 sees improvement in both ``transiting" planet and ``RV-detected" planet cases, but Kepler-21 and HD 24040 both benefit most from training on the non-simultaneous dataset. The other targets see a mix of both, and the two cases are not even consistent across targets. What patterns exist?

Kepler-21 may be the easiest case to explain. NEID RVs of Kepler-21 are less precise than the archival HARPS-N RVs, so the benefit of simultaneous NEID-TESS training is likely reduced. Additionally, archival HARPS-N RVs are simultaneous with Kepler, further reducing the importance of simultaneous NEID data. It may be that the non-simultaneous TESS sectors happened to observe the star when its activity cycle was more similar to Kepler photometry, creating a preference for that TESS training dataset.

HD 24040 is an irregular case. Our injection recovery analysis (Figure \ref{fig:HD24040}) is the most unusual of all of our targets, and we theorize that the system is not ideally suited for such. The presence of long-period, RV-detected planets seems to be interfering with the effectiveness of the Laplace Approximation, as even easily detectable injected planetary signals are not preferred in model comparison. Any conclusion taken from HD 24040 injection-recovery tests is likely suspect, and should be interpreted with caution.

HD 173701 and HD 4256 benefit the most from training on simultaneous photometry (though HD 4256 does not prefer this in the ``transiting" planet case). These systems have the longest rotation periods in our sample ( $>$ 30 days), and this may suggest some increased benefit for longer-period systems. With only six systems with which to compare results, we caution that a strong conclusion affirming this fact is not possible.

Kepler-37 sees significant improvement in the ``transiting" planet case, but in the non-transiting scenario it is similar to HD 31966 in that it sees little difference in results no matter which TESS sectors it is trained on.

Overall, it is hard to claim a clear pattern. Our comparisons are most likely hindered by the different natures of our targets: different spectral types, rotation periods, and activity levels make it difficult to identify exactly which quality is most improved, or hindered, by training on simultaneous photometry. The different qualities of the targets' archival data, too, probably affects the results, as Kepler-21, for example, has a great deal of archival HARPS-N data that are of similar precision to NEID. A future study, with a greater focus on similar stellar parameters and restricted to only the simultaneous, or near-simultaneous, RVs would likely be the best way to identify any real pattern, or lackthereof. We can conclude with confidence, however, that training data on simultaneous photometry is \textit{not} necessarily the best way to remove stellar variability from RV data. 

We conclude that if one is able to obtain enough RVs simultaneously with a photometric dataset to be able to independently recover a planetary signal, then that photometric dataset is likely the best to train upon. However, in the era of queue-scheduled RV observations, especially when utilizing data from a variety of RV instruments, one is likely better off obtaining additional RV observations in lieu of scheduling high priority time in order to achieve simultaneity. As a point of reference, the NEID queue assigns time priorities into five bins: 8$\%$ of time at priority 0, 17$\%$ at priority 1, 25$\%$ at priority 2, 25$\%$ at priority 3, and 50$\%$ at priority 4. Requesting high-priority time can seriously raise the cost of a telescope proposal, and if simultaneous RVs and photometry are unnecessary, it may not be required.

\subsection{Generalization of Results}

In \S \ref{sec:beneficial}, we discuss the implications of our analysis. These implications are dependent on our choice of activity model (a GP), and our method of translating photometric stellar activity information to RVs (detailed in \S \ref{sec:training}). Other methods may benefit more or less from simultaneous observations. Additionally, our method utilized the Laplace Approximation. While the approximation was generally valid during our analysis, a full MCMC or Nested Sampling approach might conceivably have different results.

\section{Summary}\label{sec:summary}

We perform an RV-photometry analysis of eight Kepler/K2 targets. We chose targets with known stellar activity contamination in order to study the best methods for mitigating stellar activity contamination in our RVs. We additionally obtained precise NEID observations of all targets simultaneously with TESS observations.

We then train stellar activity GP models on Kepler and TESS photometry, and use these models to recover injected planet signals for our datasets. Finally, we experiment with the limits of temporal proximity for six of our targets, and we compare the results. We conclude:

\begin{enumerate}
    \item Training stellar activity models on Kepler photometry is likely the best option when Kepler photometry is available, at least compared to TESS. Of our three Kepler targets, training on Kepler photometry improved the BF preferring injected planets by an average of 2.6 in the transiting planet case, and 3.7 in the RV-detected case.

    \item Systems with longer rotation periods (P$_{rot}$ $>$ 12 days) and higher activity levels ($\sigma_{RV}$ $>$ 5 m s$^{-1}$) may benefit more from simultaneously acquired photometric observations.

    \item Photometry simultaneous with RVs is \textit{not} necessarily the best photometric dataset to use for activity mitigation. If one is able to obtain a larger quantity of RVs that are non-simultaneous (i.e. requesting lower priority time in a queue system) than can be acquired simultaneously, this will often better constrain an RV signal.
\end{enumerate}

\section{Acknowledgements}

This paper includes data collected by the TESS mission. Funding for the TESS mission is provided by the NASA's Science Mission Directorate.

This paper includes data collected by the Kepler mission and obtained from the MAST data archive at the Space Telescope Science Institute (STScI). Funding for the Kepler mission is provided by the NASA Science Mission Directorate. STScI is operated by the Association of Universities for Research in Astronomy, Inc., under NASA contract NAS 5–26555.

Data presented were obtained by the NEID spectrograph built by Penn State University and operated at the WIYN Observatory by NOIRLab, under the NN-EXPLORE partnership of the National Aeronautics and Space Administration and the National Science Foundation. 
Based on observations at Kitt Peak National Observatory at NSF’s NOIRLab (NOIRLab Prop. ID 2021A-0396; PI: C. Beard; NOIRLab Prop. ID 2022A-377087; PI: C. Beard; NOIRLab Prop. ID 2022A-399413; PI: C. Beard;), which is managed by the Association of Universities for Research in Astronomy (AURA) under a cooperative agreement with the National Science Foundation. The authors are honored to be permitted to conduct astronomical research on Iolkam Du’ag (Kitt Peak), a mountain with particular significance to the Tohono O’odham.

We thank the NEID Queue Observers and WIYN Observing Associates for their skillful execution of our NEID observations.

This work was partially supported by NASA grant 80NSSC22K0120 to support Guest Investigator programs for TESS Cycle 4. 

This work was partially support by the Future Investigators in NASA Earth and Space Science and Technology (FINESST) program Grant No. 80NSSC22K1754.


This work utilized the infrastructure for high-performance and high-throughput computing, research data storage and analysis, and scientific software tool integration built, operated, and updated by the Research Cyberinfrastructure Center (RCIC) at the University of California, Irvine (UCI). The RCIC provides cluster-based systems, application software, and scalable storage to directly support the UCI research community. https://rcic.uci.edu


The Center for Exoplanets and Habitable Worlds is supported by the Pennsylvania State University and the Eberly College of Science

This research was carried out, in part, at the Jet Propulsion Laboratory and the California Institute of Technology under a contract with the National Aeronautics and Space Administration.


Some/all of the data presented in this article were obtained from the Mikulski Archive for Space Telescopes (MAST) at the Space Telescope Science Institute. The specific observations analyzed can be accessed via \dataset[doi: 10.17909/T98304 (All Kepler)]{https://archive.stsci.edu/doi/resolve/resolve.html?doi=10.17909/T98304}, \dataset[doi:  doi:10.17909/T9N889 (K2 C4)]{https://archive.stsci.edu/doi/resolve/resolve.html?doi=10.17909/T9N889}, \dataset[ doi:10.17909/T9TG7J (K2 C8)]{https://archive.stsci.edu/doi/resolve/resolve.html?doi=10.17909/T9TG7J}, \dataset[ doi:10.17909/T9PS3W (K2 C10)]{https://archive.stsci.edu/doi/resolve/resolve.html?doi=10.17909/T9PS3W}, \dataset[ doi:10.17909/T93M48 (K2 C13)]{https://archive.stsci.edu/doi/resolve/resolve.html?doi=10.17909/T93M48}, \dataset[ doi:10.17909/t9-z076-yp42 (K2 C17)]{https://archive.stsci.edu/doi/resolve/resolve.html?doi=10.17909/t9-z076-yp42}, and \dataset[ doi:10.17909/t9-nmc8-f686 (All TESS)]{https://archive.stsci.edu/doi/resolve/resolve.html?doi=10.17909/t9-nmc8-f686}.

\facilities{WIYN (NEID), Keck (HIRES), TNG (HARPS-N), ESO (HARPS), Lick (APF), Kepler, K2, \tess{}, Exoplanet Archive}
\software{
\texttt{astropy} \citep{astropy13,astropy18,astropy22},
\texttt{ipython} \citep{ipython07},
\texttt{lightkurve} \citep{lightkurve18},
\texttt{matplotlib} \citep{Hunter07},
\texttt{numpy} \citep{harris20},
\texttt{pandas} \citep{reback2020pandas,mckinney-proc-scipy-2010},
\texttt{RadVel}\citep{fulton18},
\texttt{scipy} \citep{2020SciPy-NMeth},
\texttt{SERVAL} \citep{zechmeister18},
}

\bibliography{bibliography}

\appendix

\renewcommand{\thefigure}{A\arabic{figure}}
\setcounter{figure}{0}
\renewcommand{\thetable}{A\arabic{table}}
\setcounter{table}{0}

We include summary plots for each of our targets.

\clearpage

\begin{figure*}
    \centering
    \includegraphics[width=\textwidth]{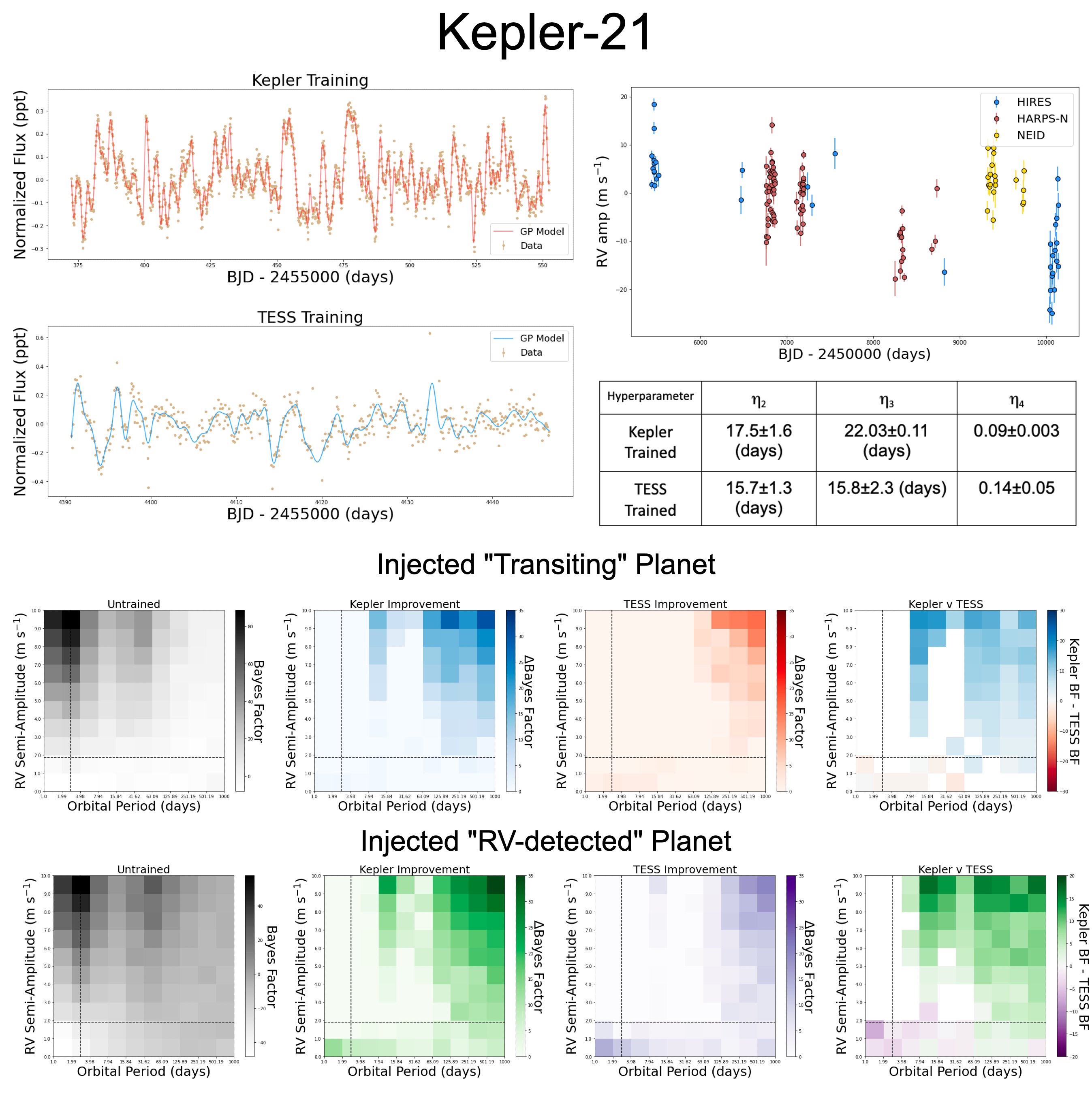}
    \caption{We include a variety of plots summarizing our training and analysis of Kepler-21. Top Left: Kepler and TESS training data, as well as our best fit GP model overlaid. Top Right: RV time series and training posteriors. Bottom: Results of our injection-recovery analysis in the two cases described in \S \ref{sec:analysis}. The left plots show the preference for models including the injected planet when no GP training is applied. The middle two plots show the improvements gained when training on Kepler or TESS. The rightmost plots highlight the differences between Kepler and TESS training. A dashed black line indicates the orbital period and amplitude of the system's known planet(s).}
    \label{fig:Kepler-21}
\end{figure*}

\begin{figure*}
    \centering
    \includegraphics[width=\textwidth]{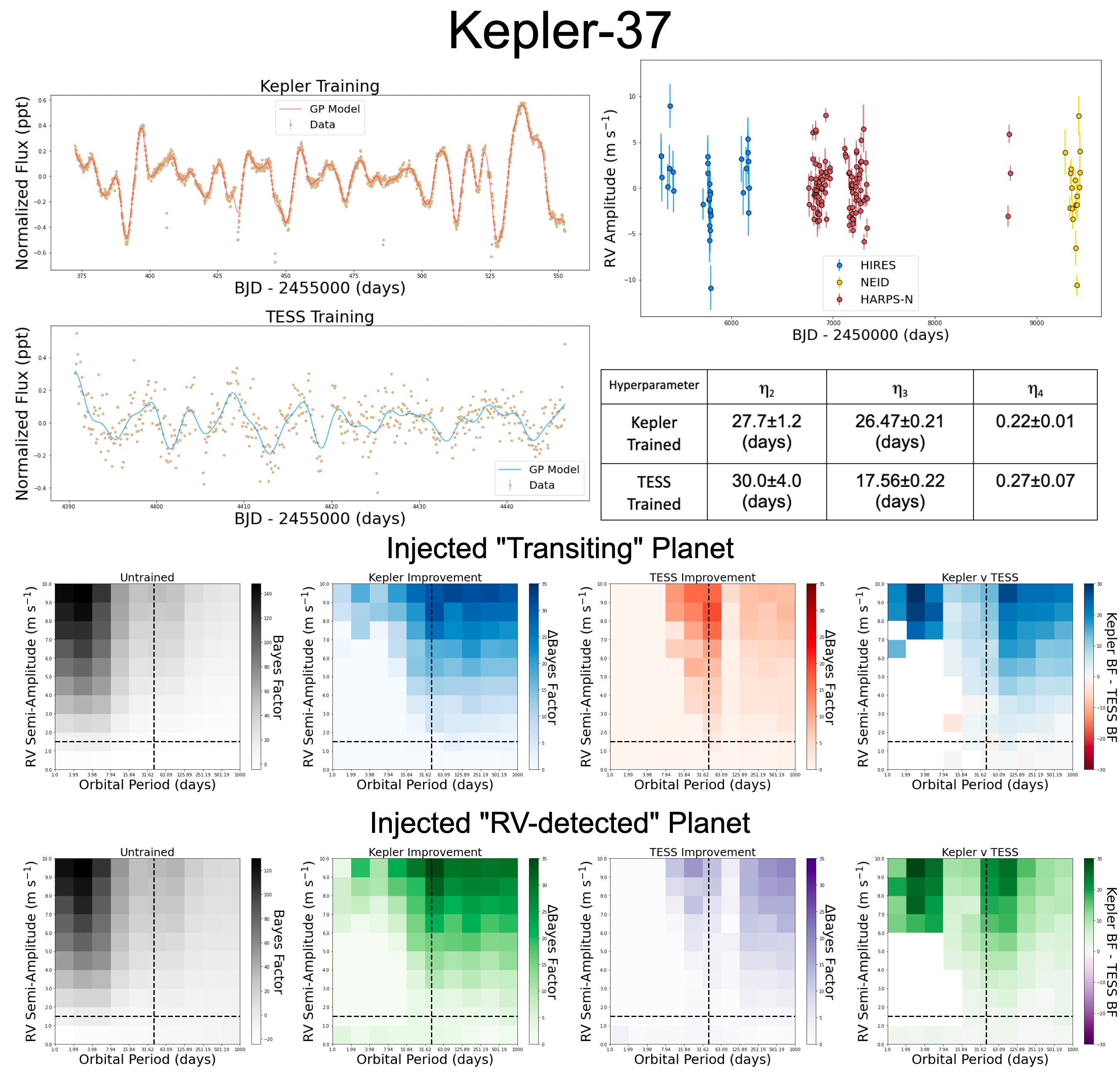}
    \caption{We include a variety of plots summarizing our training and analysis of Kepler-37. Top Left: Kepler and TESS training data, as well as our best fit GP model overlaid. Top Right: RV time series and training posteriors. Bottom: Results of our injection-recovery analysis in the two cases described in \S \ref{sec:analysis}. The left plots show the preference for models including the injected planet when no GP training is applied. The middle two plots show the improvements gained when training on Kepler or TESS. The rightmost plots highlight the differences between Kepler and TESS training. A dashed black line indicates the orbital period and amplitude of the system's known planet(s)}
    \label{fig:Kepler-37}

\end{figure*}

\begin{figure*}
    \centering
    \includegraphics[width=\textwidth]{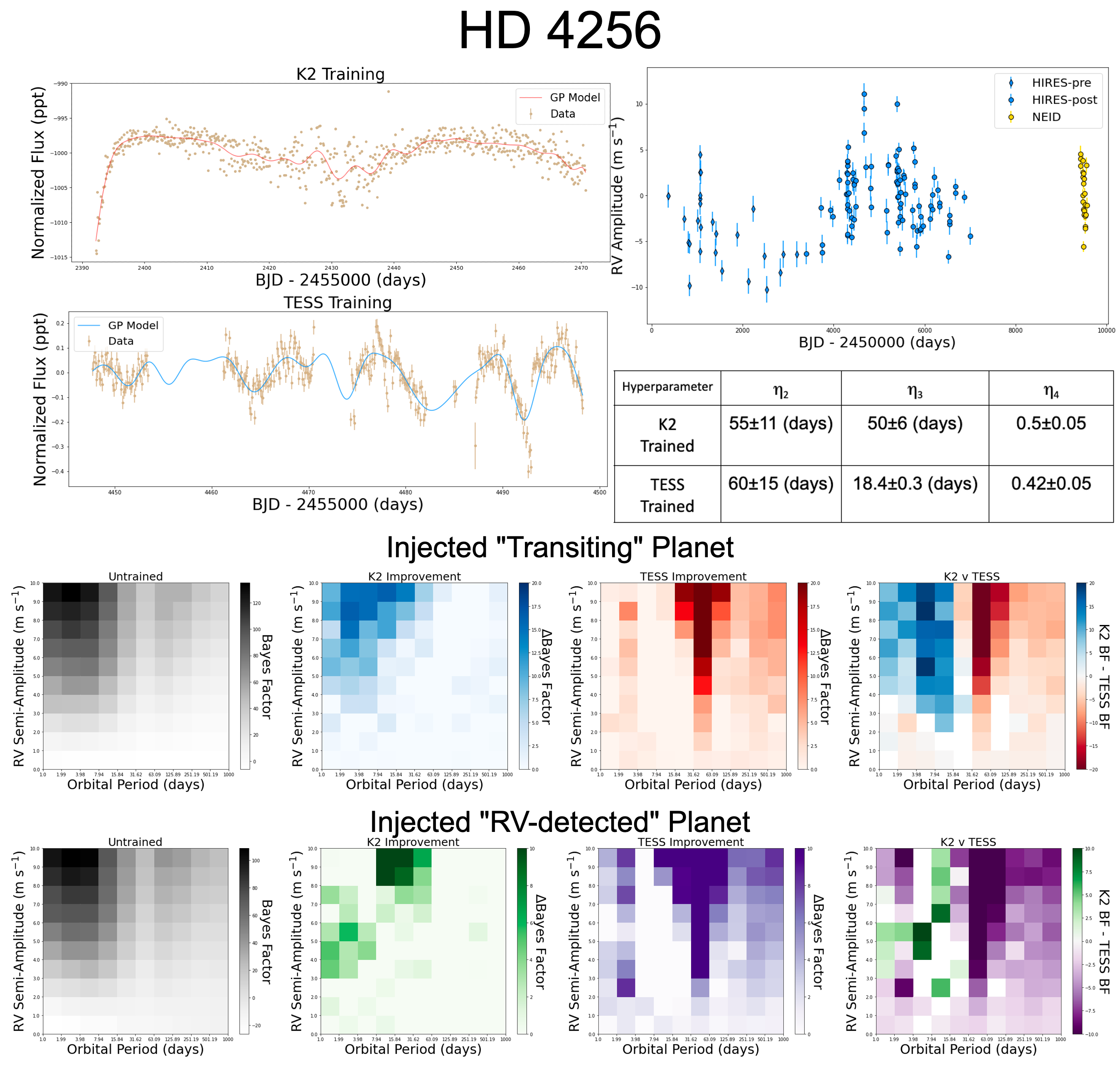}
    \caption{We include a variety of plots summarizing our training and analysis of HD 4256. Top Left: K2 and TESS training data, as well as our best fit GP model overlaid. Top Right: RV time series and training posteriors. Bottom: Results of our injection-recovery analysis in the two cases described in \S \ref{sec:analysis}. The left plots show the preference for models including the injected planet when no GP training is applied. The middle two plots show the improvements gained when training on K2 or TESS. The rightmost plots highlight the differences between Kepler and TESS training.}
    \label{fig:HD4256}

\end{figure*}

\begin{figure*}
    \centering
    \includegraphics[width=\textwidth]{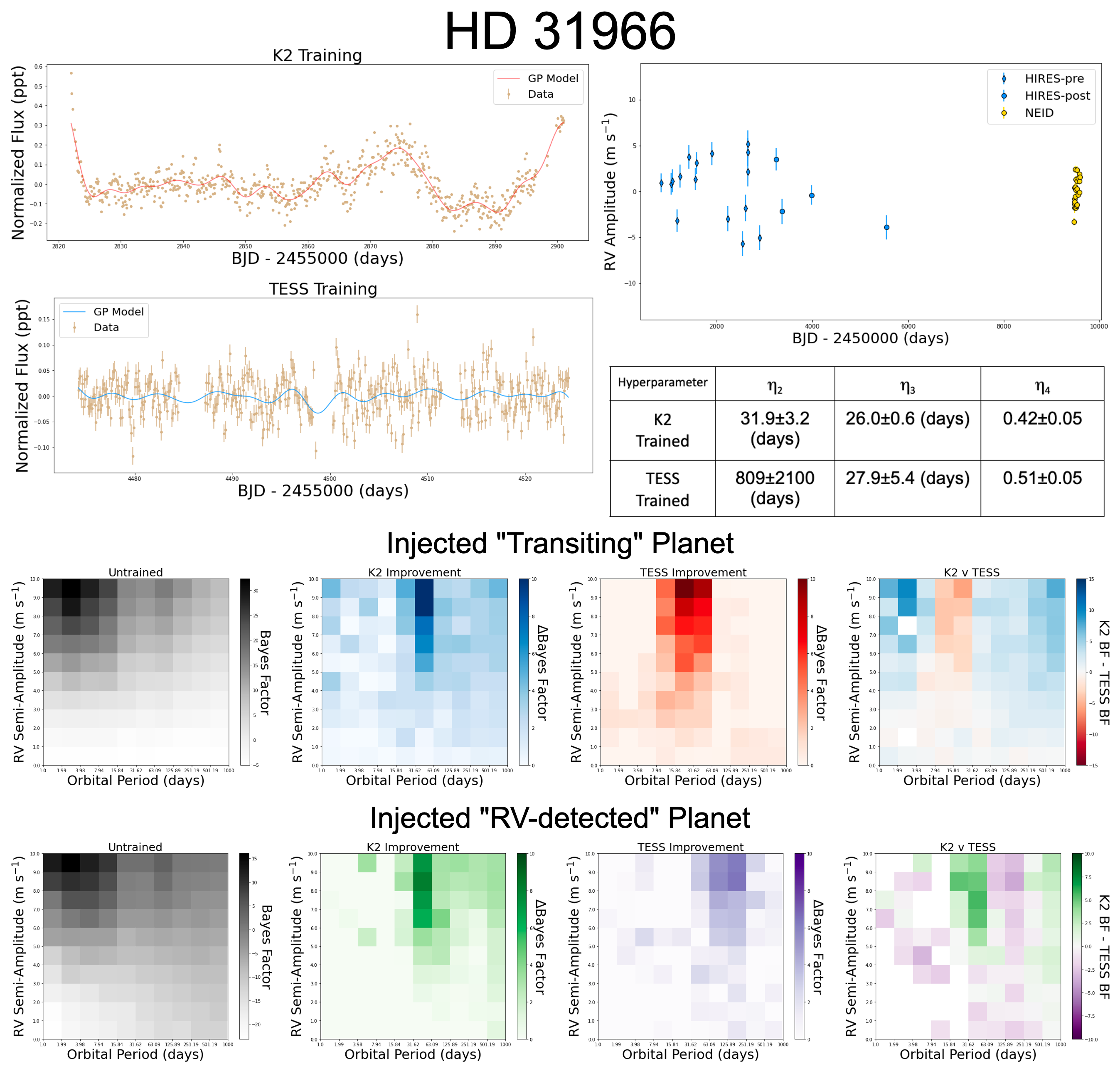}
    \caption{We include a variety of plots summarizing our training and analysis of HD 31966. Top Left: K2 and TESS training data, as well as our best fit GP model overlaid. Top Right: RV time series and training posteriors. Bottom: Results of our injection-recovery analysis in the two cases described in \S \ref{sec:analysis}. The left plots show the preference for models including the injected planet when no GP training is applied. The middle two plots show the improvements gained when training on K2 or TESS. The rightmost plots highlight the differences between Kepler and TESS training.}
    \label{fig:HD31966}
\end{figure*}

\begin{figure*}
    \centering
    \includegraphics[width=\textwidth]{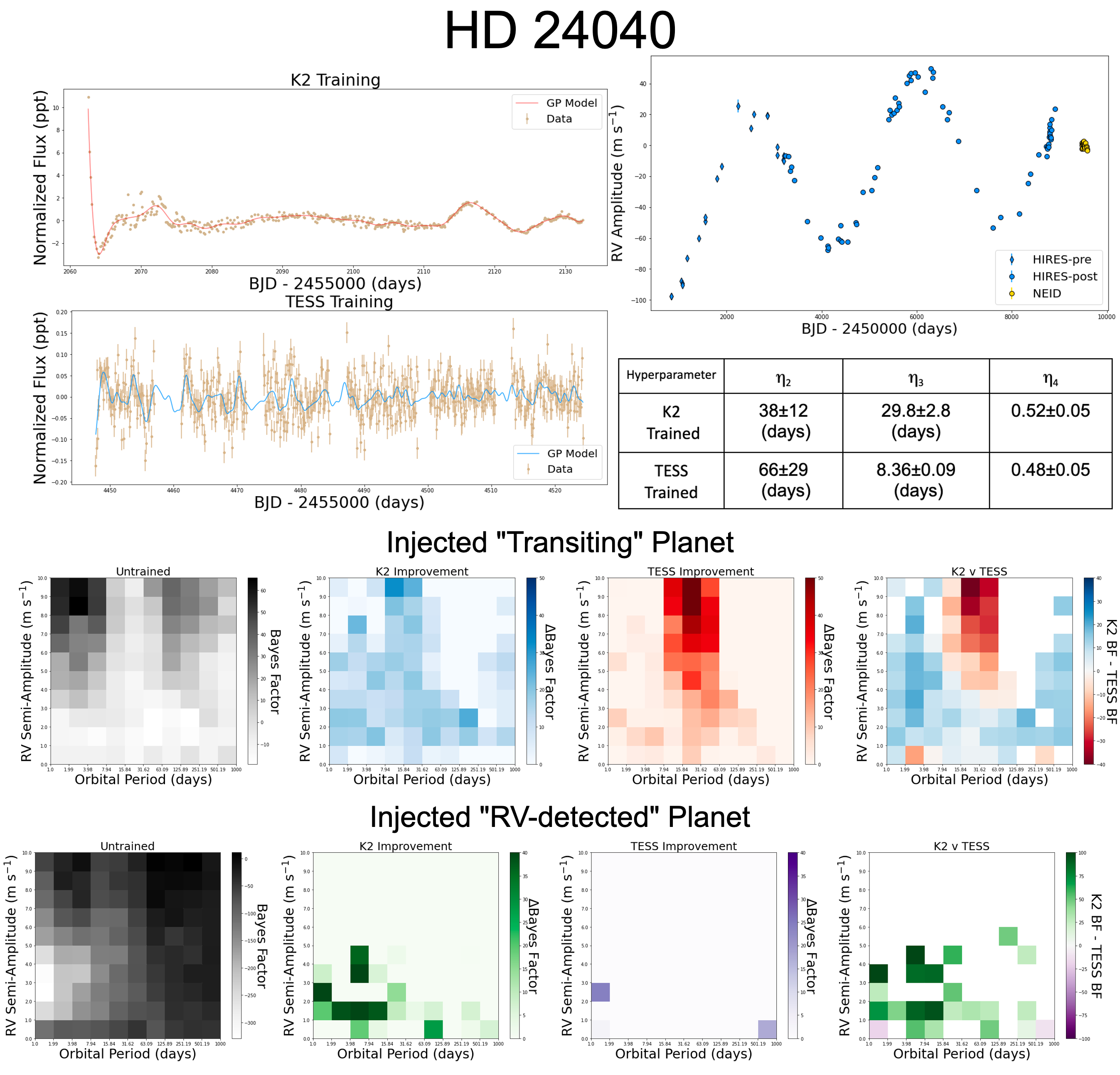}
    \caption{We include a variety of plots summarizing our training and analysis of HD 24040. Top Left: K2 and TESS training data, as well as our best fit GP model overlaid. Top Right: RV time series and training posteriors. Bottom: Results of our injection-recovery analysis in the two cases described in \S \ref{sec:analysis}. The left plots show the preference for models including the injected planet when no GP training is applied. The middle two plots show the improvements gained when training on K2 or TESS. The rightmost plots highlight the differences between Kepler and TESS training.}
    \label{fig:HD24040}
\end{figure*}

\begin{figure*}
    \centering
    \includegraphics[width=\textwidth]{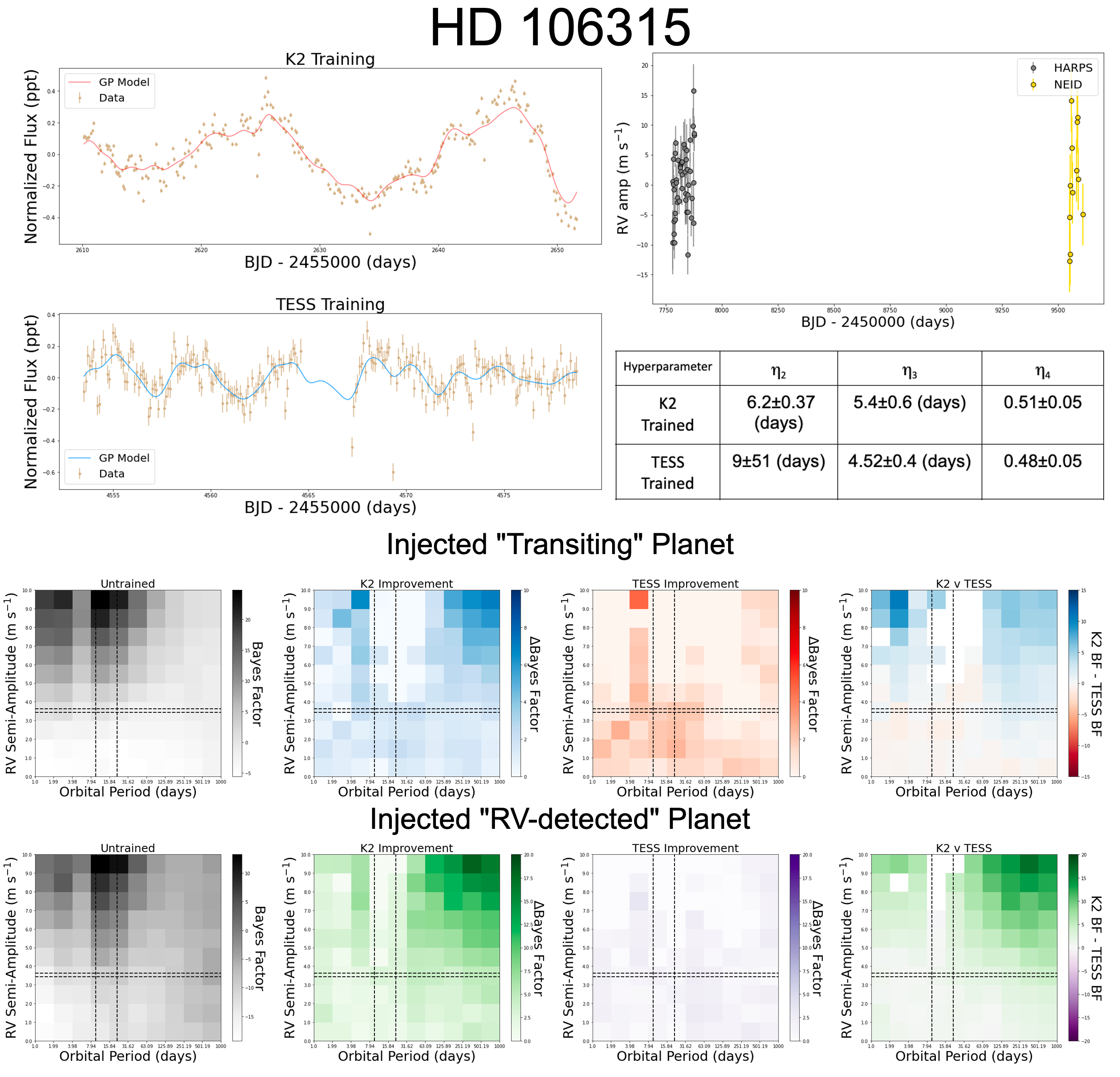}
    \caption{We include a variety of plots summarizing our training and analysis of HD 106315. Top Left: K2 and TESS training data, as well as our best fit GP model overlaid. Top Right: RV time series and training posteriors. Bottom: Results of our injection-recovery analysis in the two cases described in \S \ref{sec:analysis}. The left plots show the preference for models including the injected planet when no GP training is applied. The middle two plots show the improvements gained when training on K2 or TESS. The rightmost plots highlight the differences between Kepler and TESS training. A dashed black line indicates the orbital period and amplitude of the system's known planet(s).}
    \label{fig:HD106315}
\end{figure*}

\begin{figure*}
    \centering
    \includegraphics[width=\textwidth]{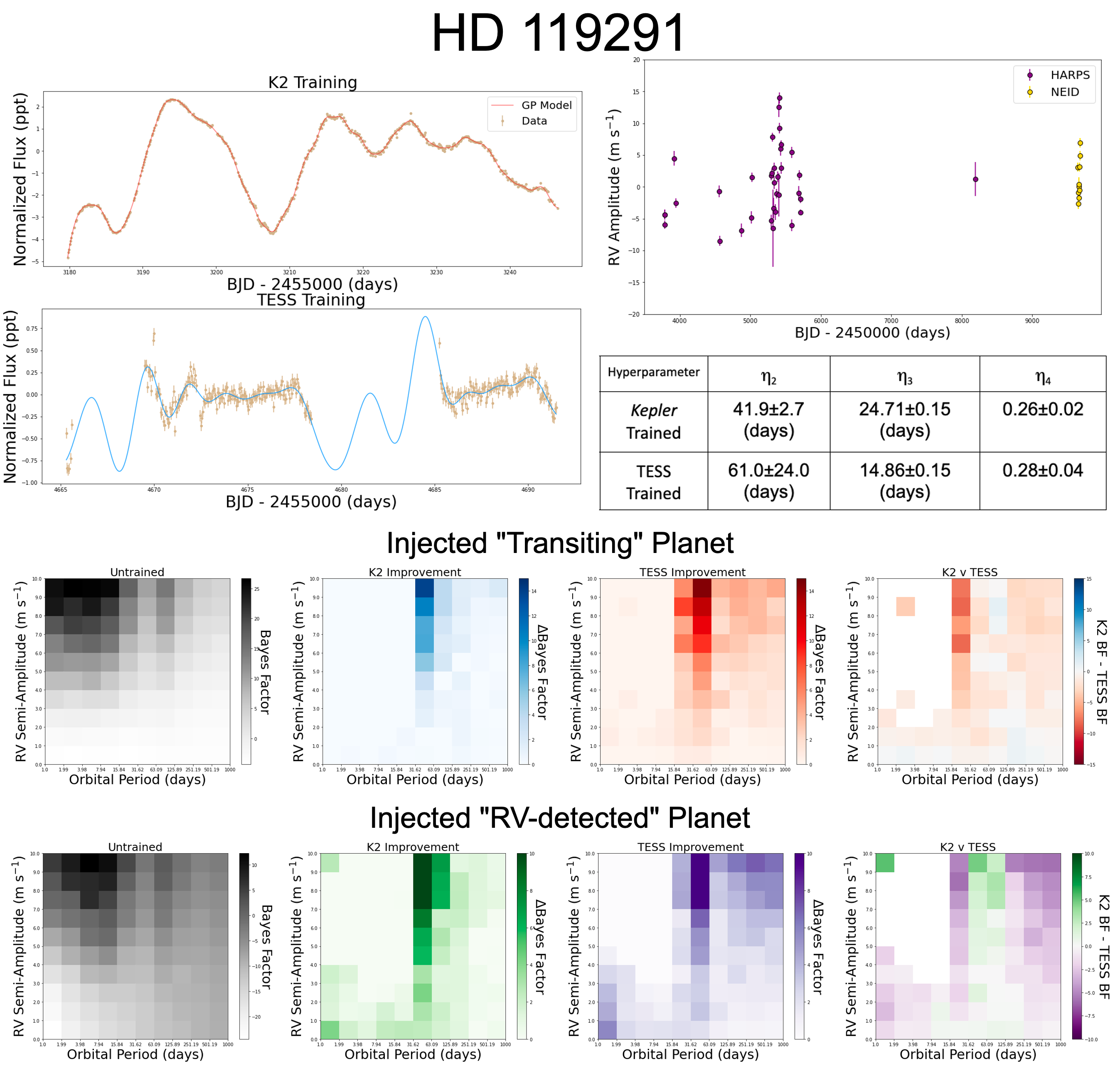}
    \caption{We include a variety of plots summarizing our training and analysis of HD 119291. Top Left: K2 and TESS training data, as well as our best fit GP model overlaid. Top Right: RV time series and training posteriors. Bottom: Results of our injection-recovery analysis in the two cases described in \S \ref{sec:analysis}. The left plots show the preference for models including the injected planet when no GP training is applied. The middle two plots show the improvements gained when training on K2 or TESS. The rightmost plots highlight the differences between Kepler and TESS training.}
    \label{fig:HD119291}
\end{figure*}

\end{document}